\documentclass{sigchi}

\nonstopmode

\setcopyright{acmlicensed}
\toappear{}

\usepackage{balance}       % to better equalize the last page
\usepackage{graphicx}      % for EPS, load graphicx instead 
\usepackage[T1]{fontenc}   % for umlauts and other diaeresis
\usepackage{txfonts}
\usepackage{mathptmx}
\usepackage[pdflang={en-US},pdftex]{hyperref}
\usepackage{xcolor,colortbl}
\usepackage{booktabs}
\usepackage{textcomp}

\usepackage{microtype}        % Improved Tracking and Kerning
\usepackage{ccicons}          % Cite your images correctly!
\usepackage{booktabs} % For formal tables
\usepackage{wrapfig}
\usepackage{subcaption}
\usepackage{listings}
\usepackage{mdframed} %nice frames
\usepackage{mathtools}
\usepackage{cuted}

\usepackage{todonotes}

\def\plaintitle{Knitting Skeletons: A Computer-Aided Design Tool\\ for Shaping and Patterning of Knitted Garments}
\def\plainauthor{Alexandre Kaspar, Liane Makatura and Wojciech Matusik}

\def\plainkeywords{
Computerized Knitting;
CAD; %Algorithmic Design;
Shaping;
Patterning}

\makeatletter
\def\url@leostyle{%
  \@ifundefined{selectfont}{
    \def\UrlFont{\sf}
  }{
    \def\UrlFont{\small\bf\ttfamily}
  }}
\makeatother
\def\pprw{8.5in}
\def\pprh{11in}

\definecolor{linkColor}{RGB}{6,125,233}
\hypersetup{%
  pdftitle={\plaintitle},
  pdfauthor={\plainauthor},
  pdfkeywords={\plainkeywords},
  pdfdisplaydoctitle=true, % For Accessibility
  bookmarksnumbered,
  pdfstartview={FitH},
  colorlinks,
  citecolor=black,
  filecolor=black,
  linkcolor=black,
  urlcolor=linkColor,
  breaklinks=true,
  hypertexnames=false
}

\newcommand{\etal}{{\em et al.}}
\newcommand{\subfig}[3]{
    \begin{subfigure}{#1\linewidth}
    \includegraphics[width=\linewidth]{#2}
    \caption{#3}
    \end{subfigure}
}
\newcommand{\subfigh}[3]{
    \subcaptionbox{#3}{
    \includegraphics[height=#1]{#2}
    }
}

\definecolor{course}{HTML}{919FFF}
\definecolor{wale}{HTML}{FF731F}

\definecolor{codegreen}{rgb}{0,0.6,0}
\definecolor{codegray}{rgb}{0.5,0.5,0.5}
\definecolor{codepurple}{rgb}{0.58,0,0.82}
\definecolor{backcolour}{rgb}{1,1,1} 
 
 \lstdefinelanguage{JavaScript}{
  keywords={typeof, new, true, false, try, finally, function, return, null, catch, switch, var, if, in, while, do, else, case, break, let, of},
  ndkeywords={class, export, boolean, throw, implements, import, this},
  sensitive=false,
  comment=[l]{//},
  morecomment=[s]{/*}{*/},
  morestring=[b]',
  morestring=[b]"
}
 
\lstdefinestyle{mystyle}{
    backgroundcolor=\color{backcolour},   
    commentstyle=\em\color{codegreen},
    keywordstyle=\bf\color{magenta},
    numberstyle=\tiny\color{codegray},
    stringstyle=\color{codepurple},
    basicstyle=\footnotesize,
    breakatwhitespace=false,         
    breaklines=true,                 
    captionpos=b,                    
    keepspaces=true,                 
    numbers=left,                    
    numbersep=10pt,                  
    showspaces=false,                
    showstringspaces=false,
    showtabs=false,                  
    tabsize=2
}
\begin{document}

\title{\plaintitle}

\numberofauthors{3}
\author{%
  \alignauthor{Alexandre Kaspar\\
    \affaddr{MIT CSAIL}\\
    \affaddr{Cambridge, MA, USA}\\
    \email{akaspar@mit.edu}}\\
  \alignauthor{Liane Makatura\\
    \affaddr{MIT CSAIL}\\
    \affaddr{Cambridge, MA, USA}\\
    \email{makatura@mit.edu}}\\
  \alignauthor{Wojciech Matusik\\
    \affaddr{MIT CSAIL}\\
    \affaddr{Cambridge, MA, USA}\\
    \email{wojciech@csail.mit.edu}}\\
}

\maketitle

% methods for figures
%
% Teaser
%

\newcommand{\figTeaser}[1][h]{
\begin{strip}
\centering
\includegraphics[width=\linewidth]{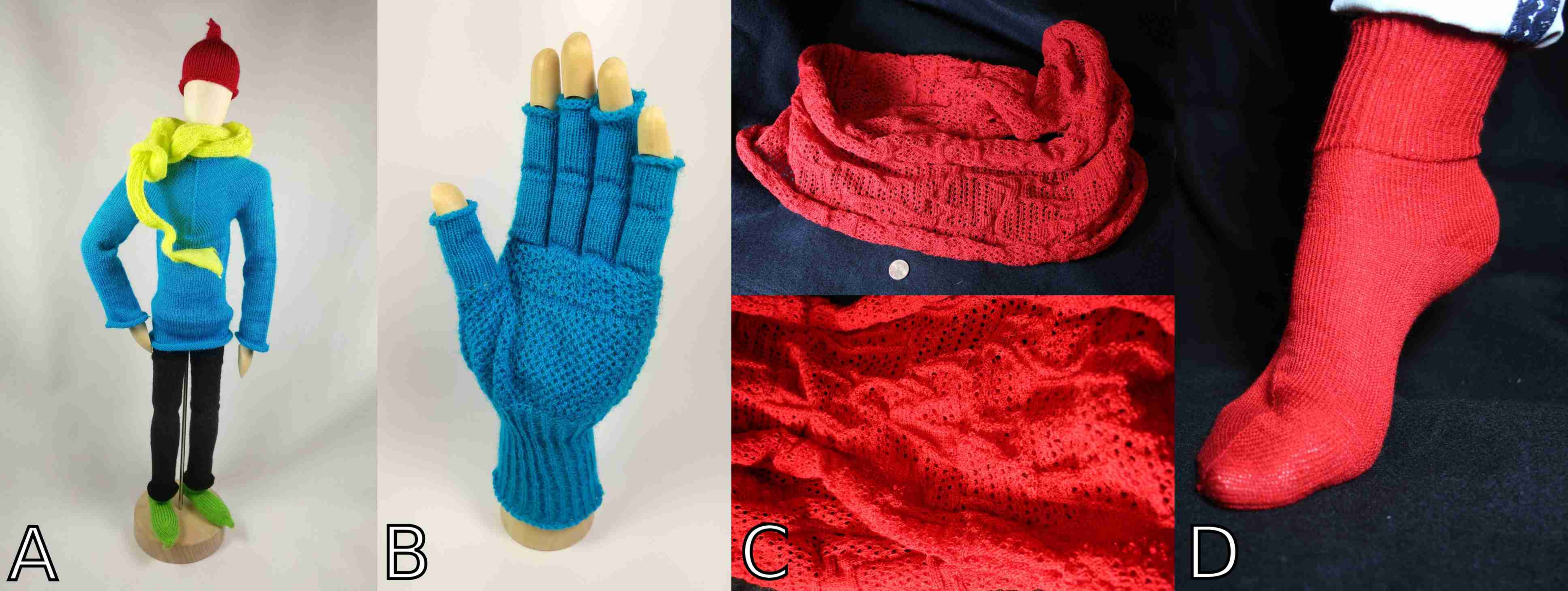}
\captionof{figure}{
Samples of knitted garments designed and customized with our tool, and fabricated on a wholegarment industrial knitting machine.
(A) Various garment prototypes on a 12 inch mannequin;
(B) a glove with lace patterns on its palms;
(C) an infinity scarf with noisy lace patterns; and
(D) a sock with a ribbed cuff.
}
\label{fig:teaser}
\end{strip}
}

%
% Needle bed figure
%

\newcommand{\figIntro}[1][t]{
\begin{figure}[#1]
\centering
\includegraphics[width=\linewidth]{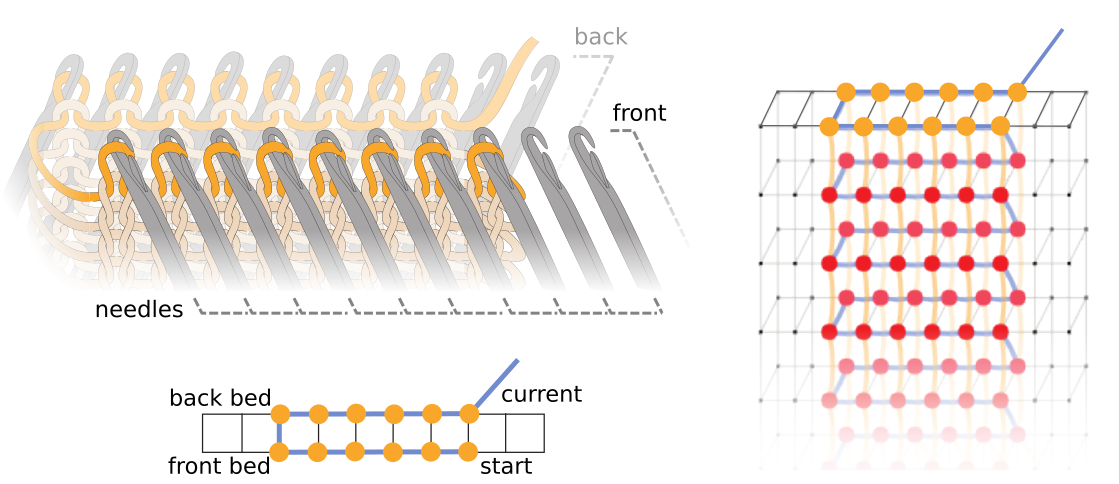}
\caption{
Illustrations of important components of a V-bed knitting machine.
\emph{Top-left}: needle beds holding a tubular structure of yarn across both sides.
\emph{Bottom-left}: a corresponding top diagram.
\emph{Right}: example of that top diagram as part of our needle bed visualization.
}
\label{fig:intro}
\end{figure}
}

%
% Stitch unit
%

\newcommand{\figStitchUnit}[1][t]{
\begin{figure}[#1]
\centering
\def\stitchHeight{2.8cm}
\includegraphics[height=\stitchHeight,trim={0.5cm 0 0.5cm 0},clip]{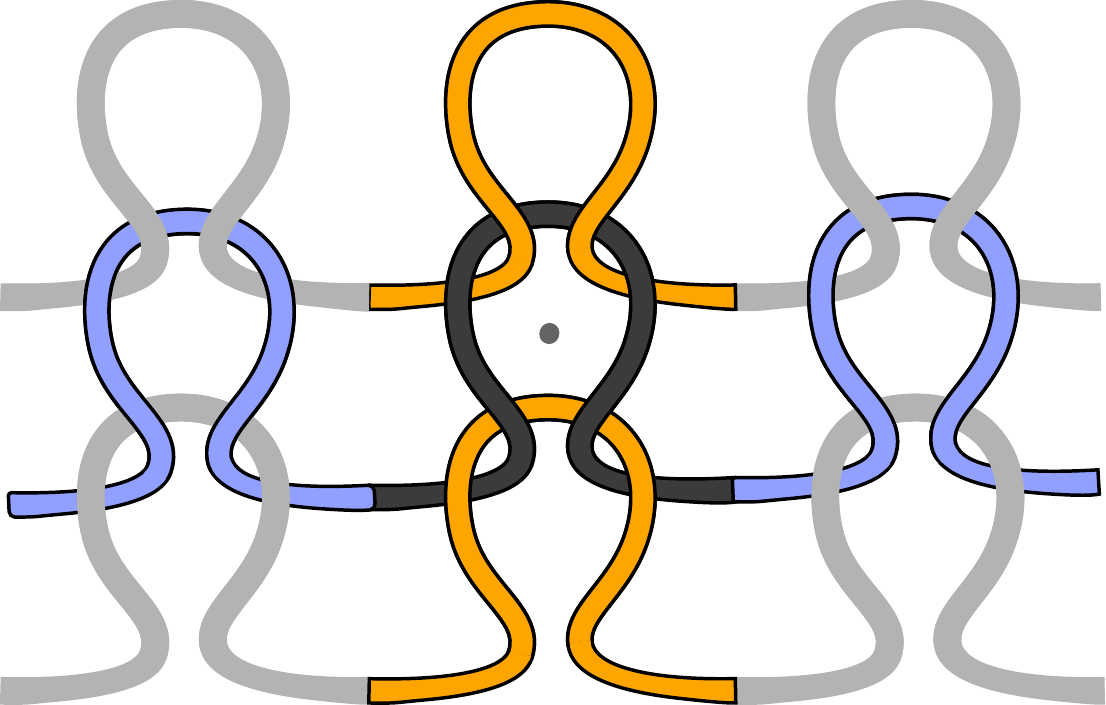}
\hfill
\includegraphics[height=\stitchHeight,trim={2cm 0 2cm 0},clip]{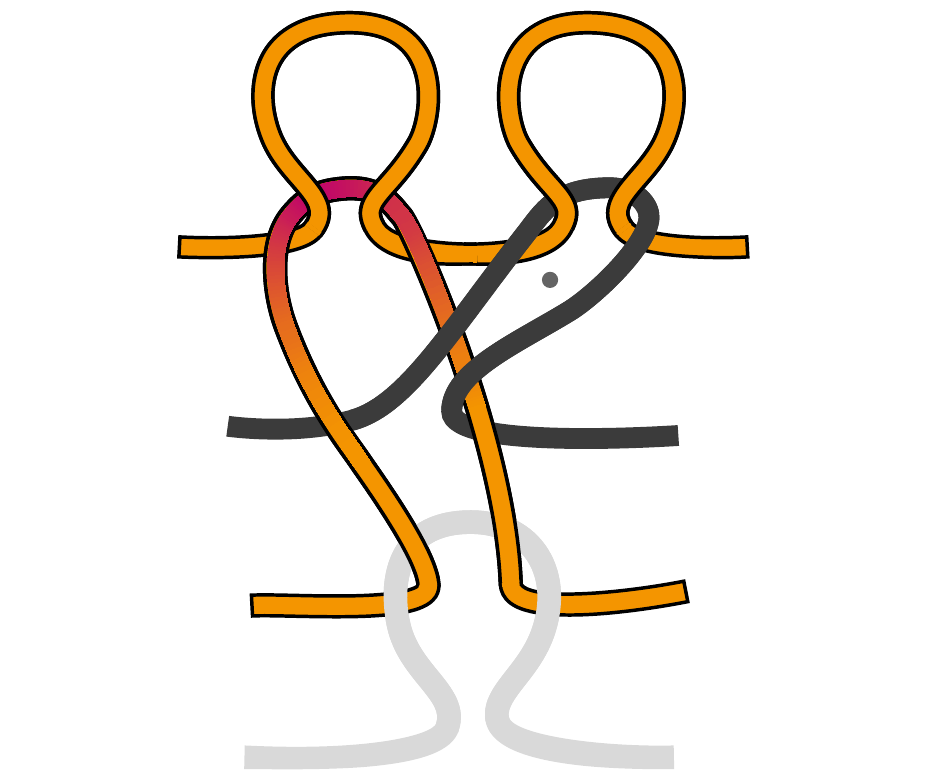}
\hfill
\includegraphics[height=\stitchHeight,trim={2cm 0 2cm 0},clip]{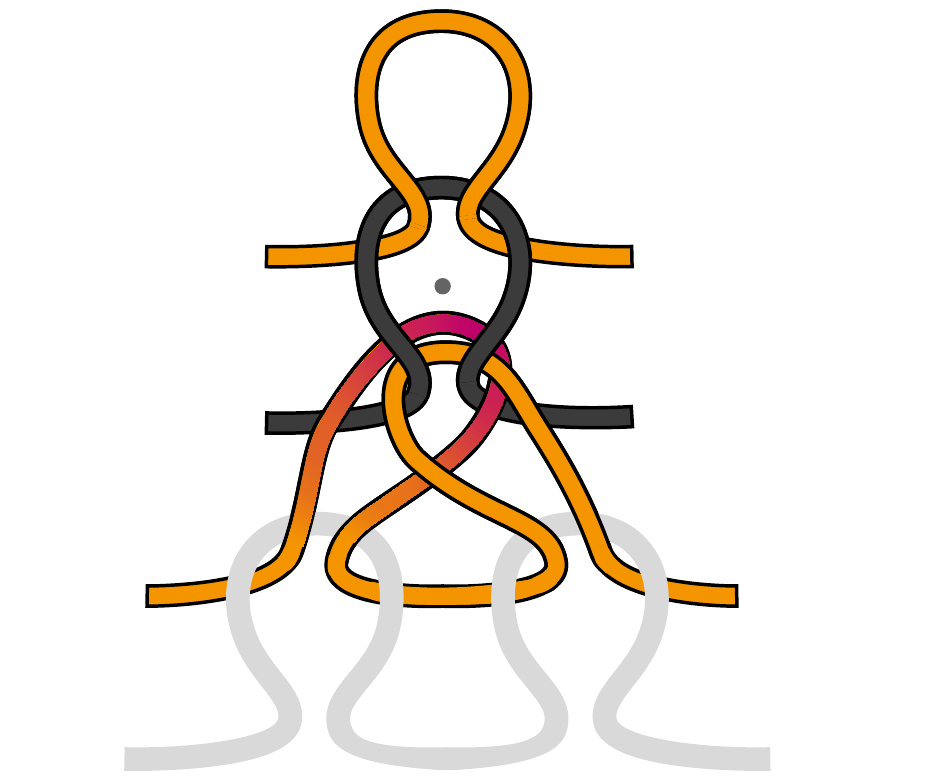}
\caption{
\emph{Left}: the base stitch unit in \textbf{black},
its {\color{course}course connections} in {\color{course}purple},
and its {\color{wale}wale ones} in {\color{wale}orange}.
One can think of courses as "rows" and wales as "columns".
\emph{Middle}:~width increase by \emph{split stitch}.
\emph{Right}:~width decrease by merging neighboring stitches using move transfers.
}
\label{fig:stitch}
\end{figure}
}

%
% Skeleton graph
%

\definecolor{sheet}{HTML}{4980F3}
\definecolor{split}{HTML}{D546D1}

\newcommand{\figSkeletonGraph}[1][t]{
\begin{figure}[#1]
\centering
\includegraphics[width=\linewidth]{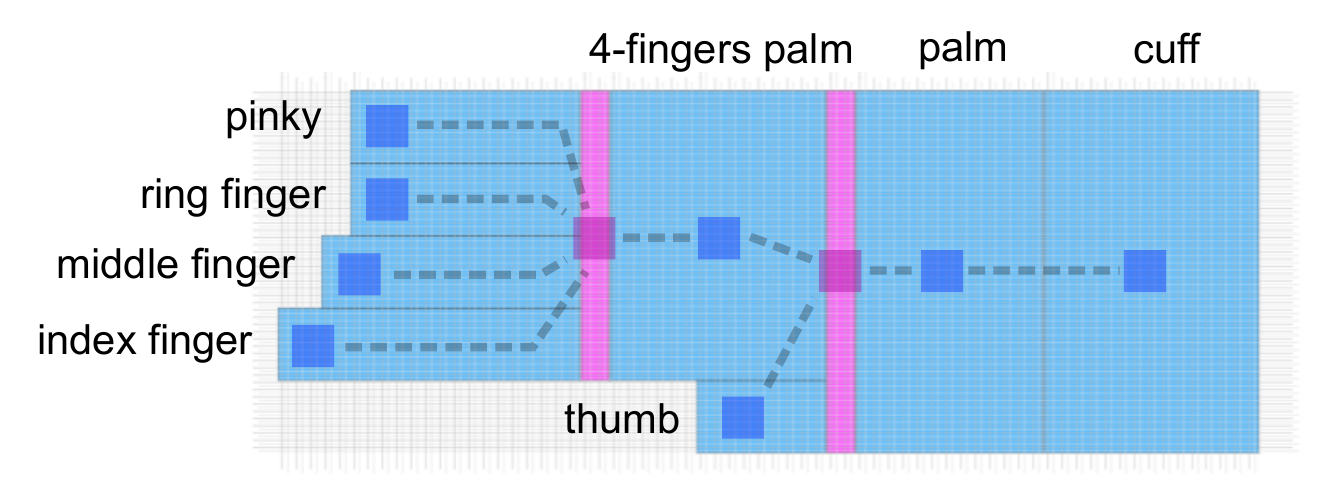}
\caption{
Sideways view of a compact glove in our system.
The underlying skeleton graph is highlighted on top, with \emph{tubular sheet} nodes in \textcolor{sheet}{blue} and \emph{split} nodes in \textcolor{split}{fuchsia}.}
\label{fig:skeleton}
\end{figure}
}

%
% Stitch graph
%

\newcommand{\figStitchGraph}[1][t]{
\begin{figure}[#1]
\centering
\includegraphics[width=0.3\linewidth]{figures/outputs/graph}
\includegraphics[width=0.3\linewidth]{figures/outputs/graph-front}
\includegraphics[width=0.3\linewidth]{figures/outputs/graph-pattern}
\caption{Simple graph visualization:
(left) the whole graph,
(center) the masked graph showing only the front side,
and (right) the graph with pattern annotations that updates
as the user modifies the pattern -- here, a 1 by 1 rib.
}
\label{fig:2d}
\end{figure}
}

%
% Time-needle bed layout
%
\newcommand{\figTNB}[1][t]{
\begin{figure*}[#1]
\centering
\subfig{0.32}{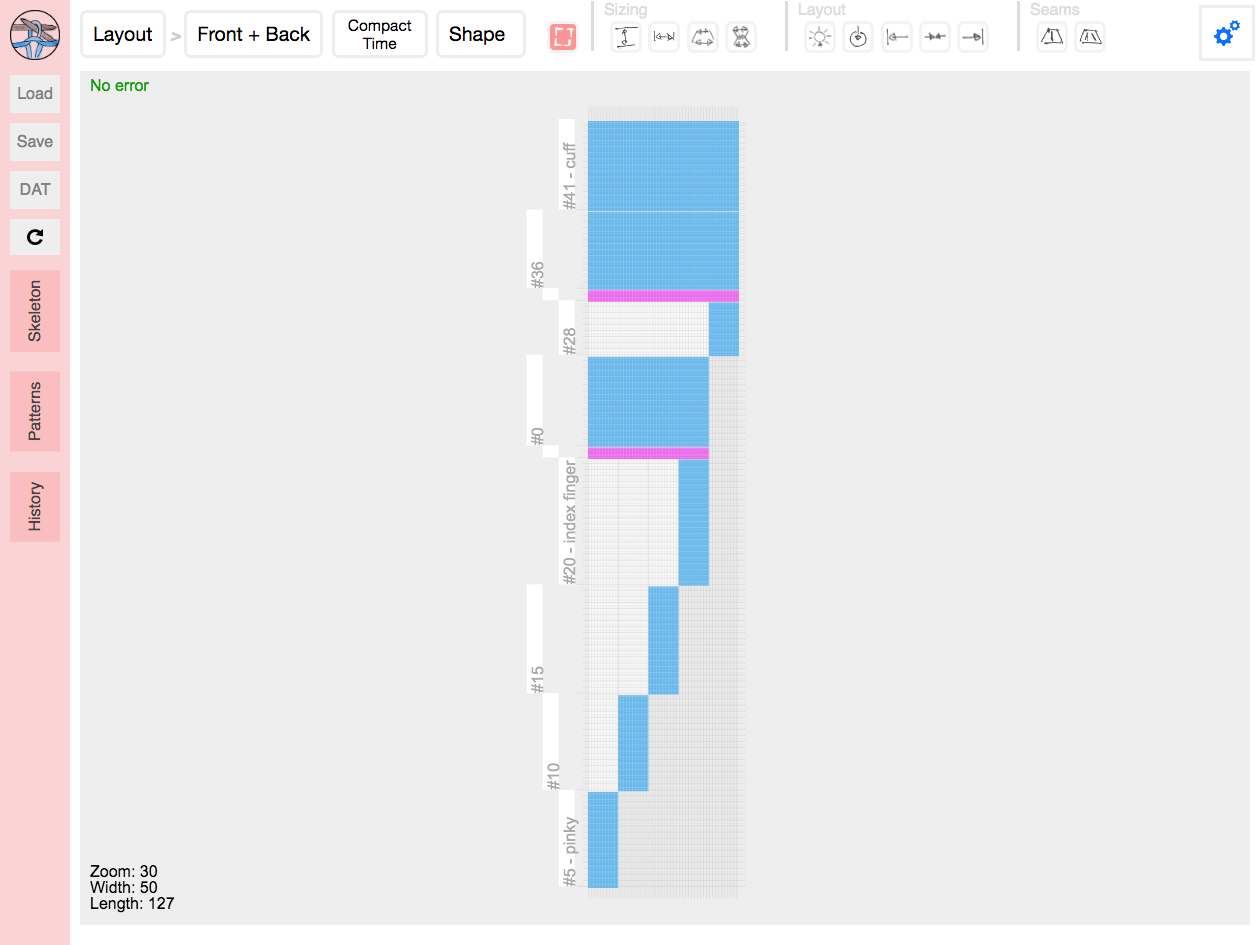}{Full time-needle bed layout}
\subfig{0.32}{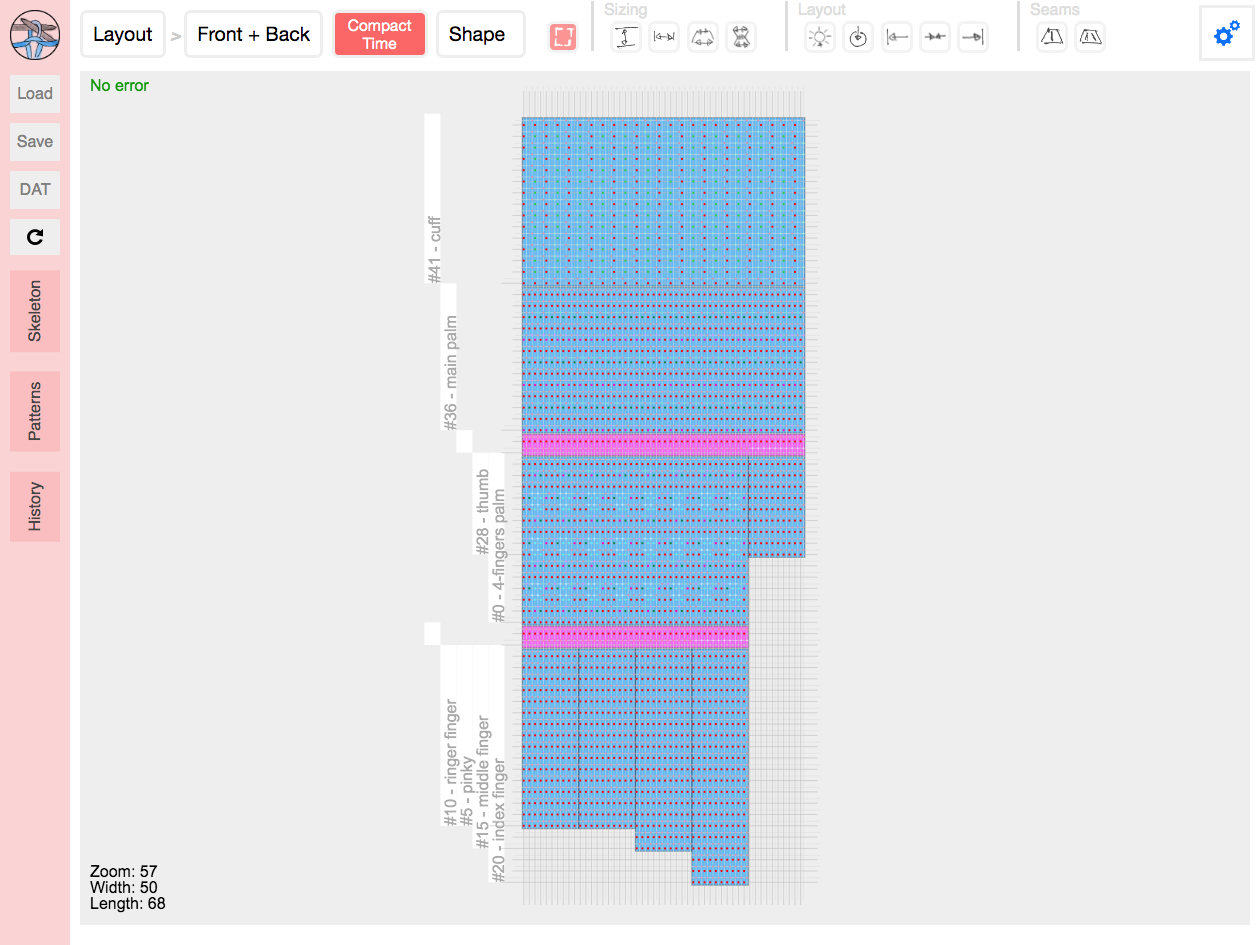}{Compacted layout for local editing}
\subfig{0.32}{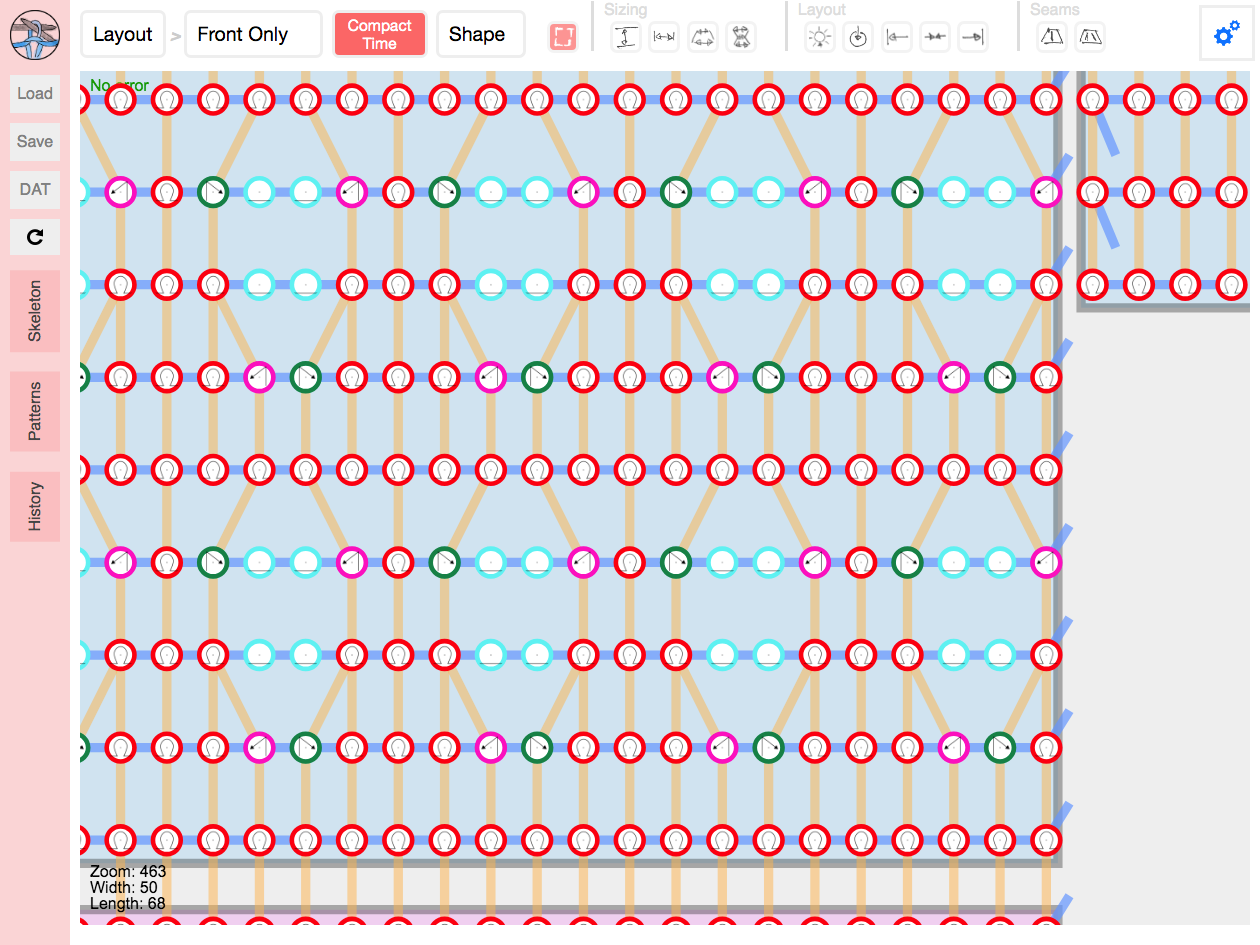}{Zoom on local yarn pattern}
\caption{
The \emph{time-needle bed} depicts the knitting process over time. We provide a \emph{compact} version that collapses suspended stitches to allow a local composition of primitives instead of the traditional composition over time.
By zooming on the layout, we can inspect the local patterning operations and the simulated pattern flow.
Finally, the user can view either side of the garment.
}
\label{fig:tnb}
\end{figure*}
}

%
% Interpretable warnings
%
\newcommand{\figWarnings}[1][t]{
\begin{figure}[#1]
\centering
\begin{minipage}{0.49\linewidth}
\includegraphics[width=\linewidth]{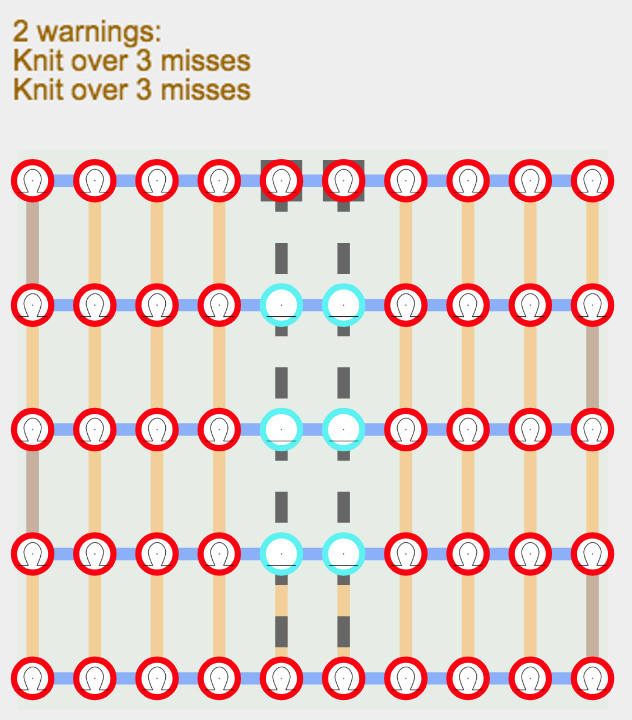}
\end{minipage}
\hfill
\begin{minipage}{0.49\linewidth}
\includegraphics[width=\linewidth]{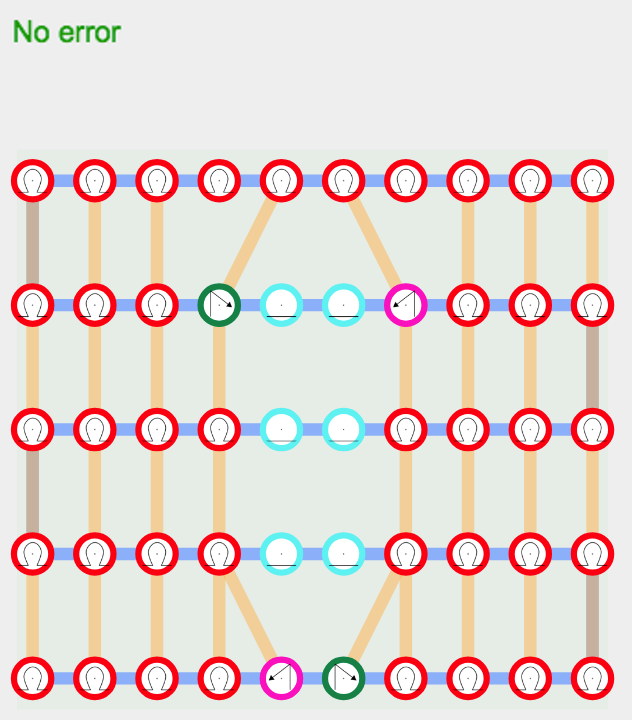}
\end{minipage}
\caption{Warnings regarding a long-term dependency that would collapse the yarn (left).
By highlighting the conflict dependencies, the user can more easily fix the pattern (right).
}
\label{fig:warnings}
\end{figure}
}

%
% Force-simulation
%
\newcommand{\figSimulation}[1][t]{
\begin{figure}[#1]
\centering
\includegraphics[width=\linewidth]{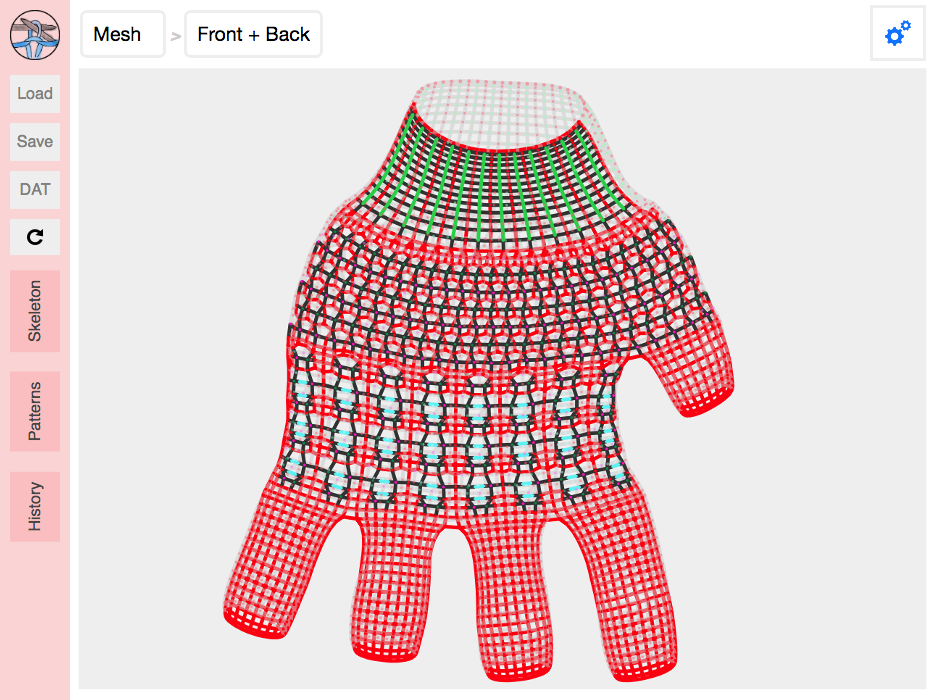}
\caption{Force-layout simulation to preview the impact of the yarn stress forces on the final shape.}
\label{fig:simulation}
\end{figure}
}

%
% Low-level instructions
%
\newcommand{\figMachineCode}[1][t]{
\begin{figure}[#1]
\centering
\includegraphics[width=0.9\linewidth]{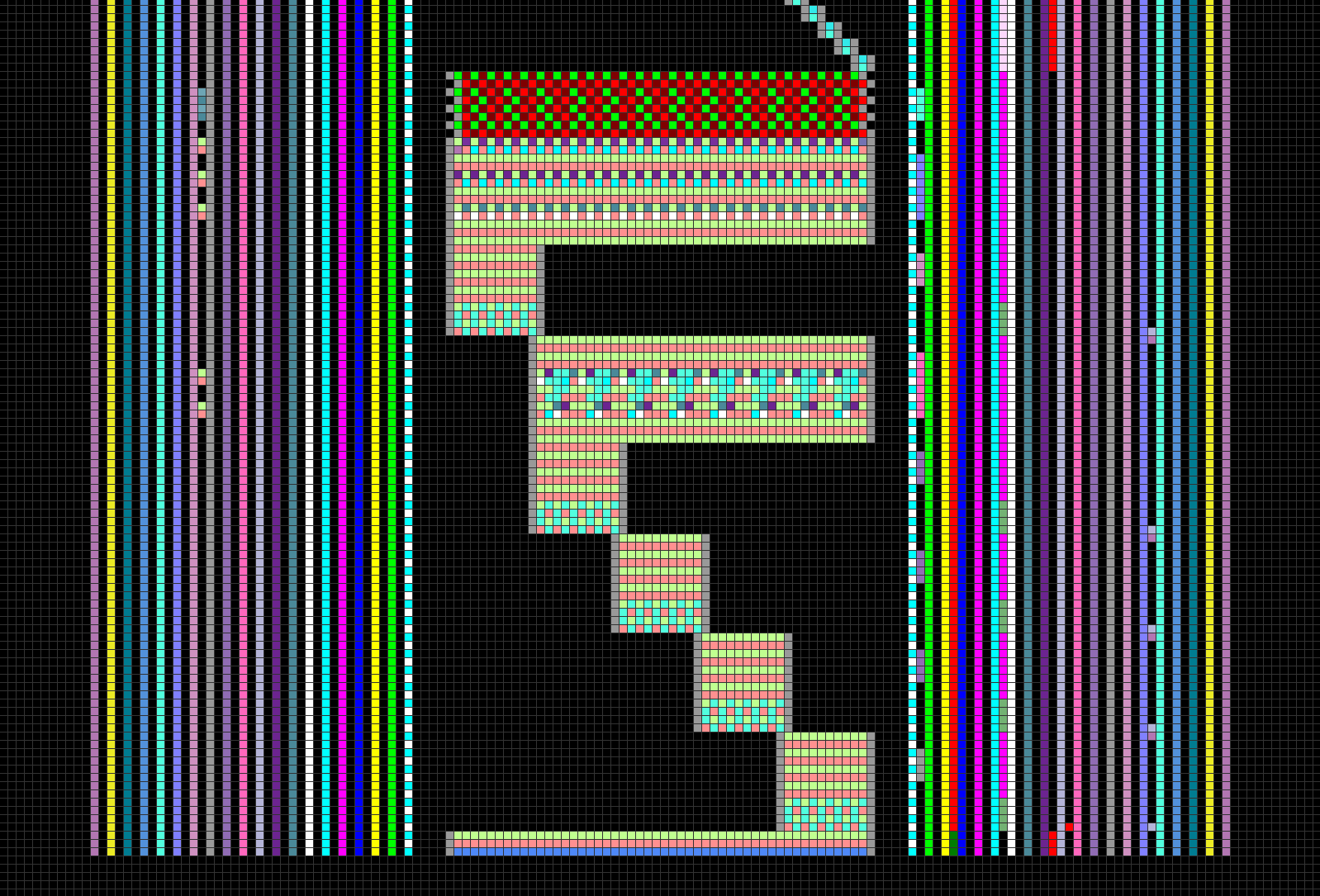}
\caption{
Part of the low-level instructions for a simplified version of the glove, to be processed with KnitPaint~\protect\cite{Shima11}.
The x axis corresponds to the needle bed.
The y axis corresponds to time.
The center shows individual machine instructions, and the sides include various options.
} 
\label{fig:machinecode}
\end{figure}
}

%
% Sheet parameters and illustration
%

\newcommand{\figSheet}[1][t]{
\begin{figure}[#1]
\centering
\begin{subfigure}[c]{0.45\linewidth}
  \includegraphics[width=\linewidth]{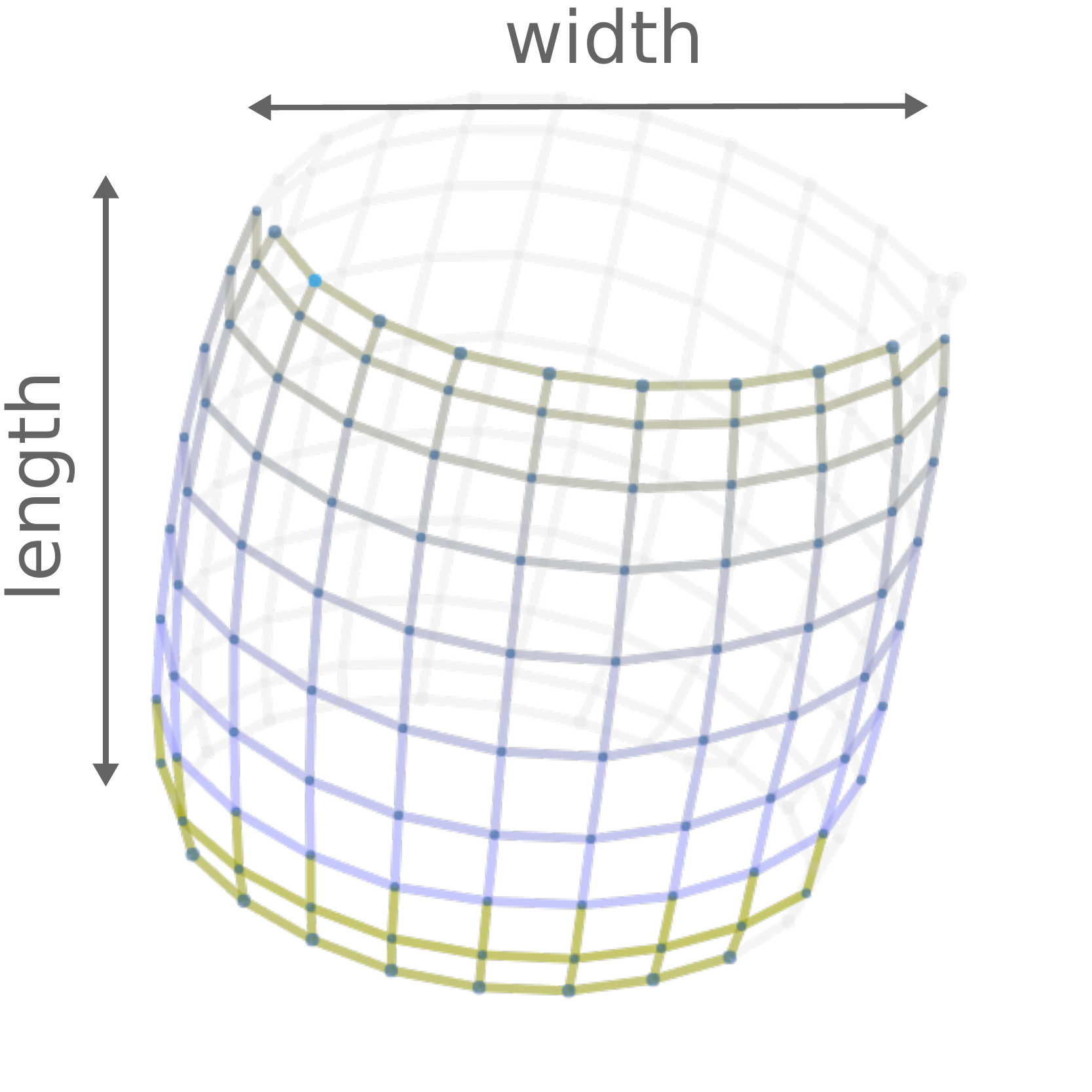}
\end{subfigure}
\begin{subfigure}[c]{0.5\linewidth}
\begin{tabular}{lp{2.2cm}}
\toprule
\bf Property & \bf Values \\
\midrule
Type & \texttt{Flat}, \texttt{Tubular} \\
Length & Integer \\
Width & $[0;1] \to \mathbb{R}_{>0}$ \\
Shaping & \small\tt Uniform, Sides, Left, Right, Center, Custom \\
Alignment & \tt Left, Right, Center\\
\midrule
Bottom & Interface \\
Top & Interface \\
\bottomrule
\end{tabular}
\end{subfigure}
\caption{A tubular sheet, and the table of its properties}
\end{figure}
}

%
% Joint parameters and illustration
%

\newcommand{\figJoint}[1][t]{
\begin{figure}[#1]
\begin{subfigure}[c]{0.45\linewidth}
  \includegraphics[width=\linewidth]{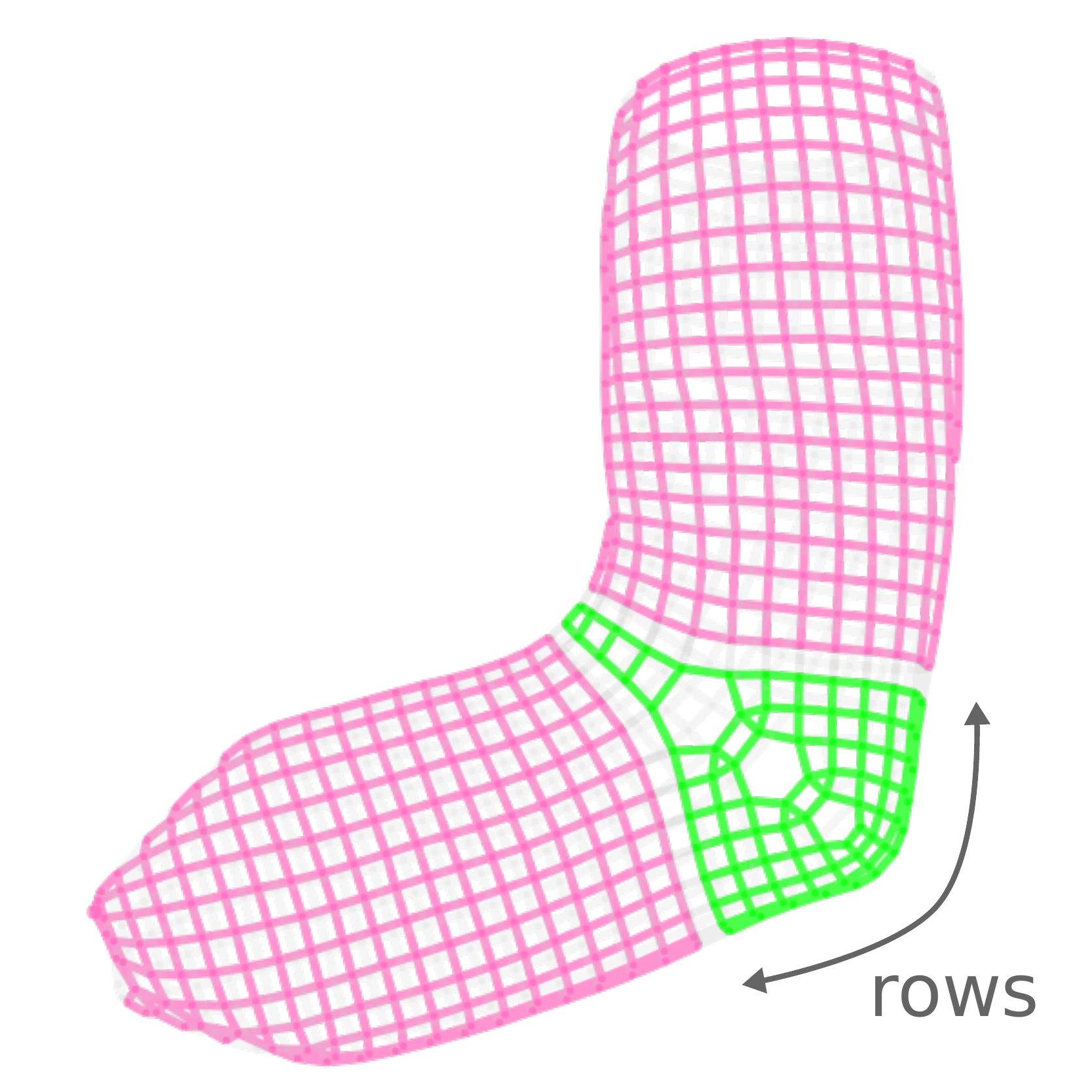}
\end{subfigure}
\begin{subfigure}[c]{0.5\linewidth}
\begin{tabular}{l p{2.2cm}}
\toprule
\bf Property & \bf Values \\
\midrule
Rows & Integer \\
Width & $[0;1] \to \mathbb{R}_{>0}$ \\
Layout & $[0;1]$ or "auto" \\
Alignment & \tt Left, Right, Center \\
\midrule
Bottom & Interface \\
Top & Interface \\
\bottomrule
\end{tabular}
\end{subfigure}
\caption{
Joint primitive as the heel of a sock,
and the table of its properties}
\end{figure}
}

%
% Split parameters and illustration
%

\newcommand{\figSplit}[1][t]{
\begin{figure}[#1]
\begin{subfigure}[c]{0.45\linewidth}
  \includegraphics[width=\linewidth]{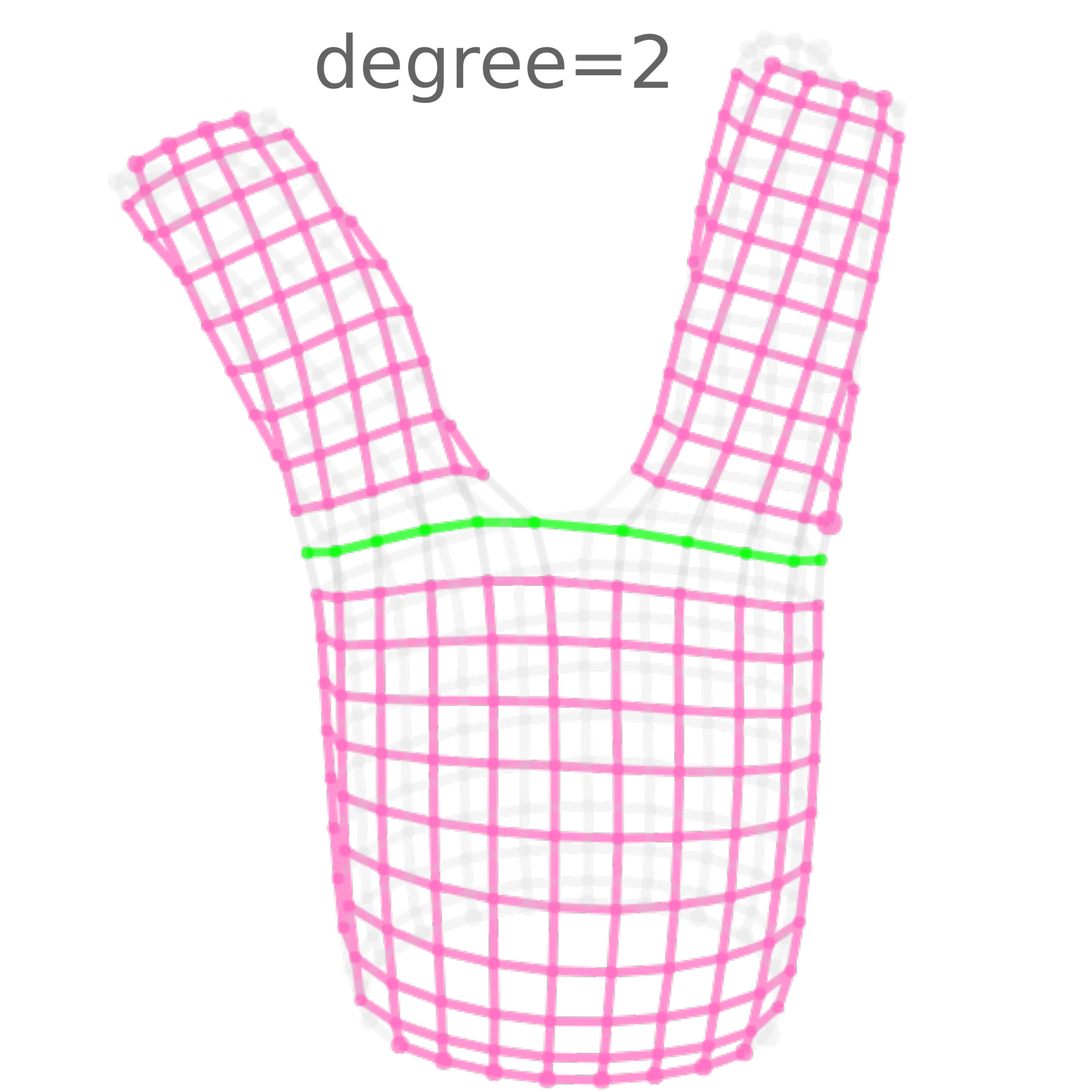}
\end{subfigure}
\begin{subfigure}[c]{0.5\linewidth}
\begin{tabular}{l p{2.2cm}}
\toprule
\bf Property & \bf Values \\
\midrule
Degree & Integer \\
Layout & $[0;1]^d$ or "auto" \\
Alignment & \tt Uniform, Left, Right, Center \\
\midrule
Folded & \texttt{True} or \texttt{False} \\
\midrule
Base & Interface \\
Branches & List[Interface] \\
\bottomrule
\end{tabular}
\end{subfigure}
\caption{Split primitive between one sheet branching into two, and the table of its properties}
\end{figure}
}

\newcommand{\figFolded}[1][t]{
\begin{figure}[#1]
\centering
%\subfig{0.47}{figures/primitives/split_folded.pdf}{Tubular branches}
%\subfig{0.47}{figures/primitives/split_unfolded.pdf}{Flat branches}
%\subfig{0.47}{figures/primitives/split_folded.pdf}{\centering Tubular branches\hfill Folded = \texttt{True}}
%\subfig{0.47}{figures/primitives/split_unfolded.pdf}{\centering Flat branches\hfill Folded = \texttt{False}}
\begin{subfigure}{0.47\linewidth}
    \includegraphics[width=\linewidth]{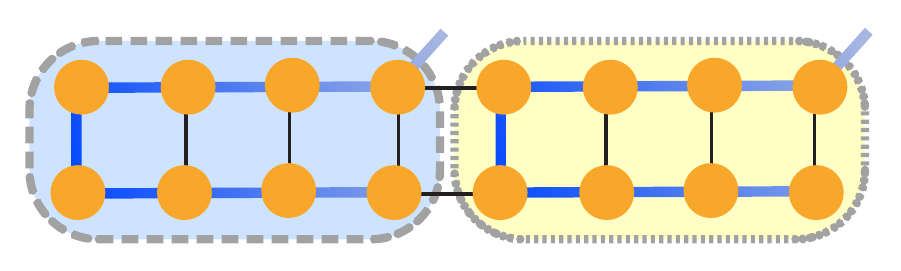}
    % @see https://tex.stackexchange.com/questions/101595/how-to-add-line-break-to-caption-without-using-caption-package
    \caption[]{\centering Tubular branches\hspace{\textwidth} Folded = \texttt{True}}
\end{subfigure}
\begin{subfigure}{0.47\linewidth}
    \includegraphics[width=\linewidth]{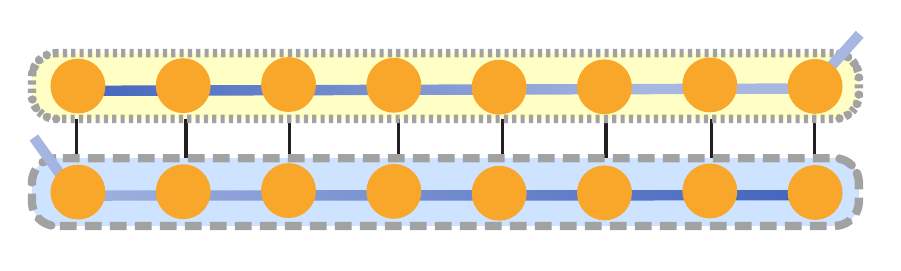}
    \caption[]{\centering Flat branches\hspace{\textwidth} Folded = \texttt{False}}
\end{subfigure}
%\begin{minipage}{0.47\linewidth}
%    \centering
%    \includegraphics[width=\linewidth]{figures/primitives/split_folded.pdf}
%    Tubular branches\\
%    Folded = \texttt{True}
%\end{minipage}
%\begin{minipage}{0.47\linewidth}
%    \centering
%   \includegraphics[width=\linewidth]{figures/primitives/split_unfolded.pdf}
%    Flat branches\\
%    Folded = \texttt{False}
%\end{minipage}
\caption{
Diagram illustrating the difference between \emph{folded} and \emph{non-folded} splits for a tubular base across the two needle beds.
The two branches are highlighted with different colors.
}
\label{fig:folded}
\end{figure}
}

%
% Shaper programs
%

\newcommand{\figShapers}[1][t]{
\begin{figure*}[#1]
\centering
\begin{subfigure}[t]{0.49\textwidth}
\includegraphics[width=\linewidth]{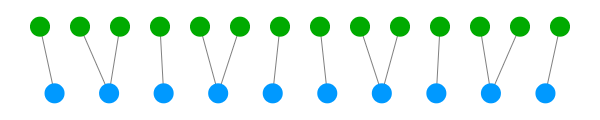}
\lstinputlisting[language=Javascript,caption=Uniform shaper program]{results/uniform.shaper}
\includegraphics[width=0.49\linewidth]{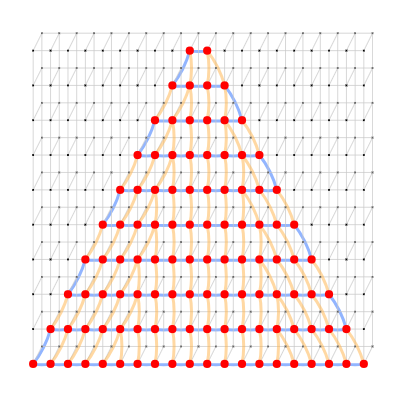}
\hfill
\includegraphics[width=0.49\linewidth,height=4cm]{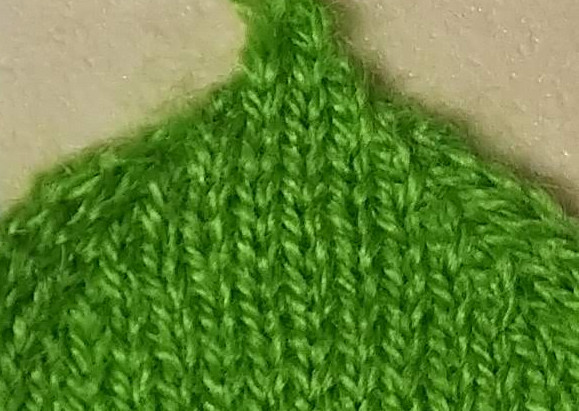}
\end{subfigure}
\begin{subfigure}[t]{0.49\textwidth}
\includegraphics[width=\linewidth]{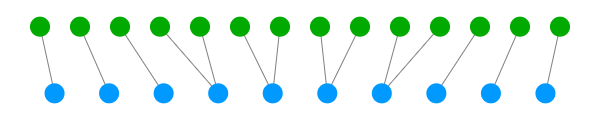}
\lstinputlisting[language=Javascript,caption=Center shaper program]{results/center.shaper}
\includegraphics[width=0.49\linewidth]{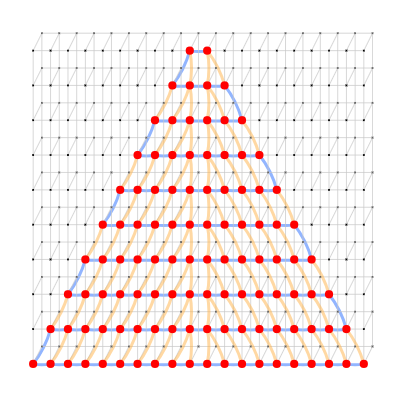}
\hfill
\includegraphics[width=0.49\linewidth,height=4cm]{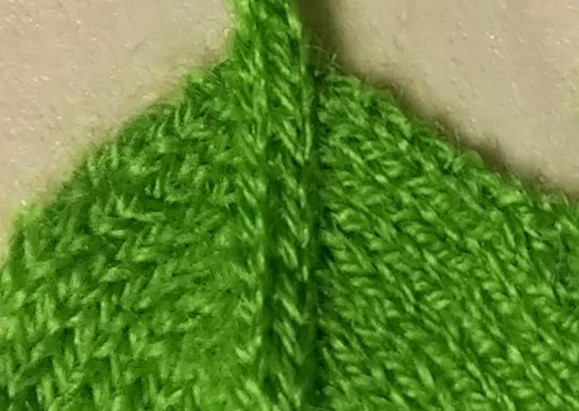}
\end{subfigure}
\caption{
Standard shaper programs:
(left) \emph{uniform} distributes the increases and decreases uniformly,
(right) \emph{center} accumulates them in the center of the course.
Notice the visible seam in the center.
At the top are program illustrations for $M=10$, $N=14$ (these numbers are here for illustration, they vary for every row in a real knitting structure).
At the bottom are the bed layouts as well as the knitted results for a flat triangular sheet.
}
\label{fig:shapers}
\end{figure*}
}

%
% Table of query commands
%

\newcommand{\tableQuery}[1][t]{
\begin{table*}[#1]
\centering
  \begin{tabular}{r p{0.45\textwidth} p{0.35\textwidth}}
  \toprule
  \bf Category & \bf Methods & \bf Explanation \\
  \midrule
  Filtering & \tt all(), filter(pred) & Stitches that match a logic predicate \\
  Sets & \tt or(x,y), and(x,y), minus(x,y), inverse() & Standard operations on sets of stitches \\
  Indexed & \tt wales(rng), courses(rng), select(c, w) & Indexed stitches within a range \\
  Neighborhood & \tt neighbors(rng), boundaries() & Stitches at some distance from selection \\
  Named & \tt named(), shape(name), itf(name) & Stitches located by a named entity \\
  Masking & \tt stretch(grid), tile(grid), img(src) & Stitches that match a given grid mask \\ 
  \bottomrule
  \end{tabular}
\caption{Our categories of pattern queries with their main methods and usage explanation}
  \label{table:query}
\end{table*}
}

%
% Figure of query illustrations
%
\newcommand{\queryz}[2]{
\begin{subfigure}[t]{0.3\linewidth}
\centering
\includegraphics[width=\linewidth]{figures/queries/#1}
#2
\end{subfigure}
}
\newcommand{\figQuery}[1][t]{
\begin{figure}[t]
\centering
\queryz{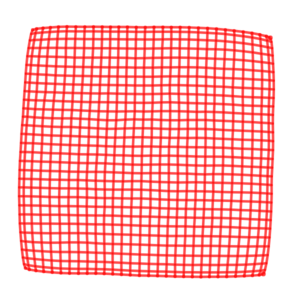}{Initial}
\queryz{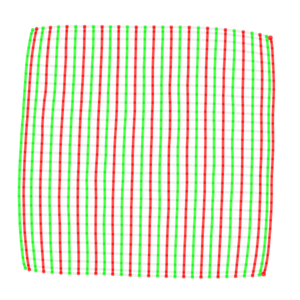}{Range of \texttt{wales}}
\queryz{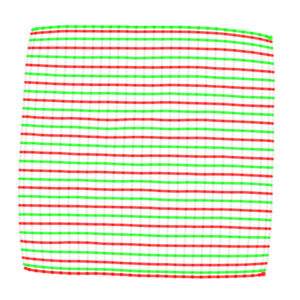}{\dots of \texttt{courses}}
\queryz{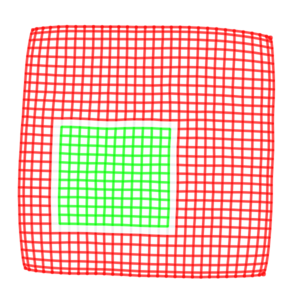}{Area \texttt{select}}
\queryz{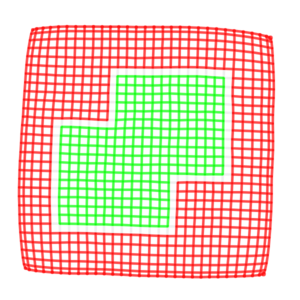}{Union} % \texttt{or}}
\queryz{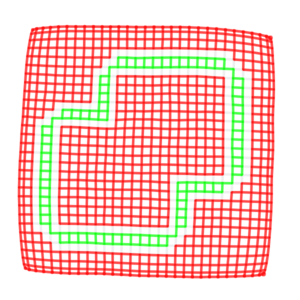}{\texttt{neighbor}}
\queryz{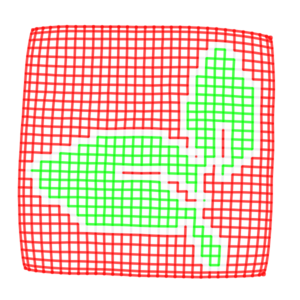}{\texttt{img} mask}
\queryz{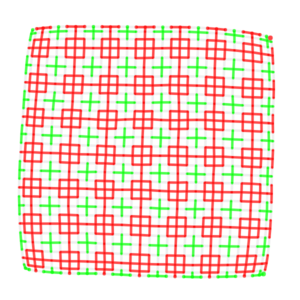}{\texttt{tile} mask}
\queryz{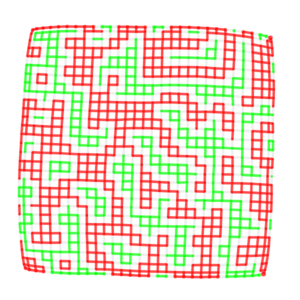}{Filter with noise}
\caption{Illustrations of some of the main pattern queries, each highlighted on a $30\times 30$ flat sheet.
}
\label{fig:query}
\end{figure}
}

%
% Illustrations of pattern types
%

\newcommand{\stitchTypesShort}[2]{
  \begin{subfigure}[t]{0.29\linewidth}
  \begin{minipage}[t][2cm][t]{\linewidth}
  	\includegraphics[width=\linewidth,height=2cm]{figures/stitches/#1.jpg}
  \end{minipage}
  \includegraphics[width=\linewidth]{figures/stitches/#1_diagram}
  \caption{#2
    \raisebox{-.3\height}{\includegraphics[width=0.5cm]{figures/stitches/#1.png}}
  }
  \end{subfigure}
}
\newcommand{\figTypesShort}[1][t]{
\begin{figure}[#1]
\centering
\stitchTypesShort{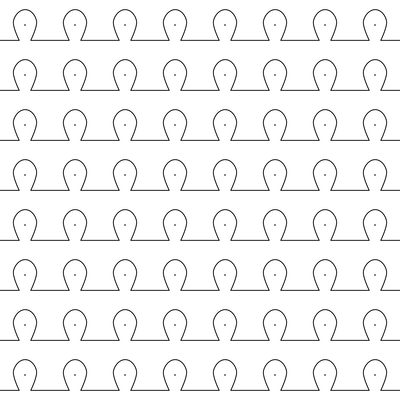}{Knit}
\stitchTypesShort{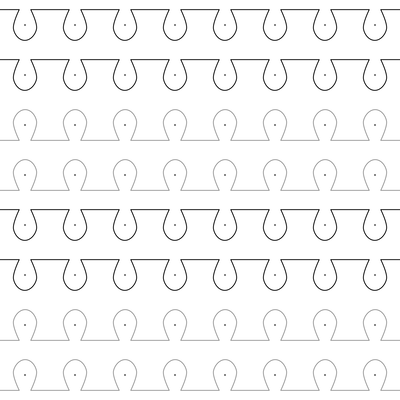}{Purl}
\stitchTypesShort{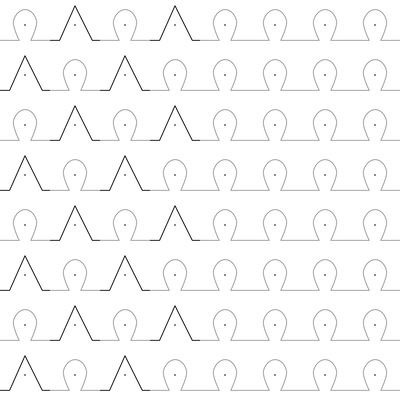}{Tuck}
\vspace{2mm}\\
\stitchTypesShort{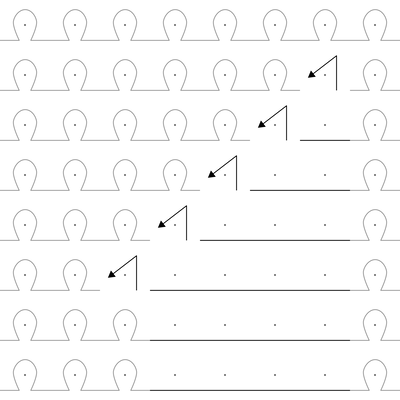}{Miss}
\stitchTypesShort{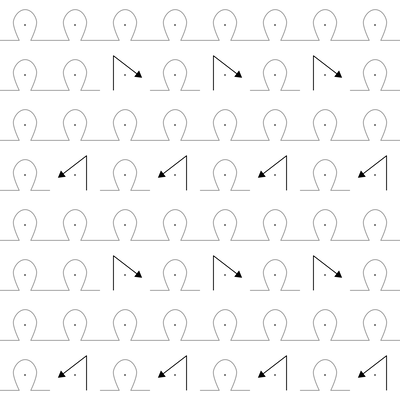}{Move}
\stitchTypesShort{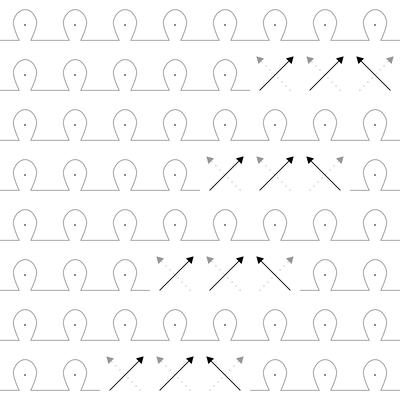}{Cross}
\caption{The different stitch operations with $8\times 8$ pattern illustrations,
both as a diagram and a knitted artifact.}
\label{fig:types}
\end{figure}
}

%
% Gauge parameter
%

\newcommand{\figGauge}[1][t]{
\begin{figure}[#1]
\centering
\begin{subfigure}{0.45\linewidth}
    \includegraphics[width=\linewidth]{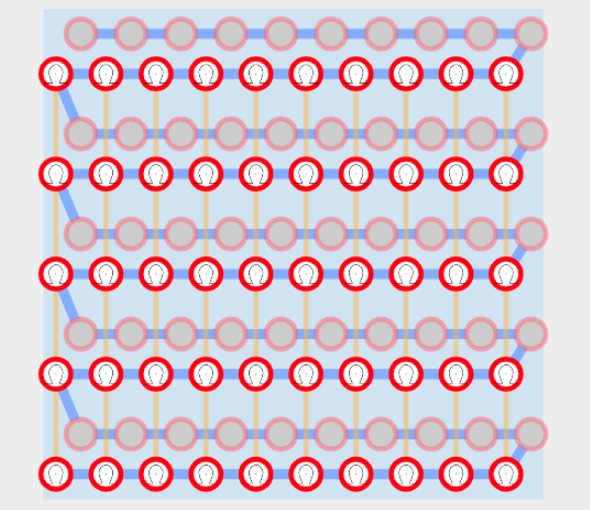}
    \caption[]{\centering Tubular in full gauge.\hfill $\textrm{width}=10$}
    %\caption{Tubular in full gauge. $\textrm{width}=10$}
\end{subfigure}
\hfill
\begin{subfigure}{0.45\linewidth}
    \includegraphics[width=\linewidth]{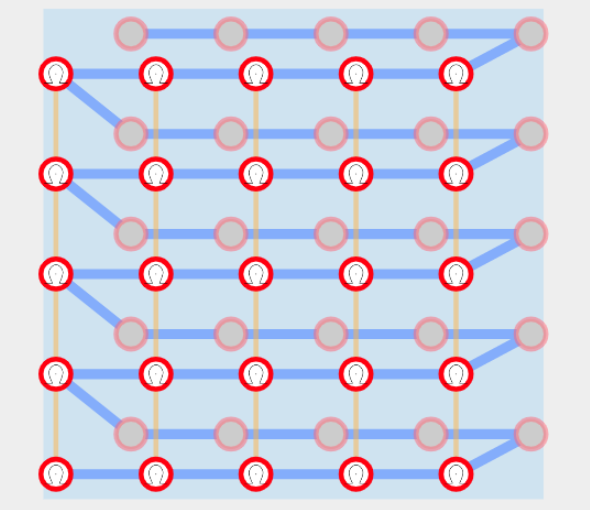}
    \caption[]{\centering Tubular in half gauge.\hfill $\textrm{width}=5$}
\end{subfigure}
\caption{
Illustration of the gauge parameter. The \texttt{width} is modified to keep the same bed support. The half-gauge variant uses different offsets between beds to allow reverse stitches.
}
\label{fig:gauge}
\end{figure}
}

%
% Layer behaviours
%

\newcommand{\figBehaviours}{
\begin{figure*}[t]
\centering
\newcommand{\layerHeight}{2.1cm}
\subfigh{\layerHeight}{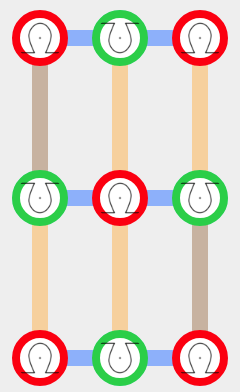}{Pattern}
\subfigh{\layerHeight}{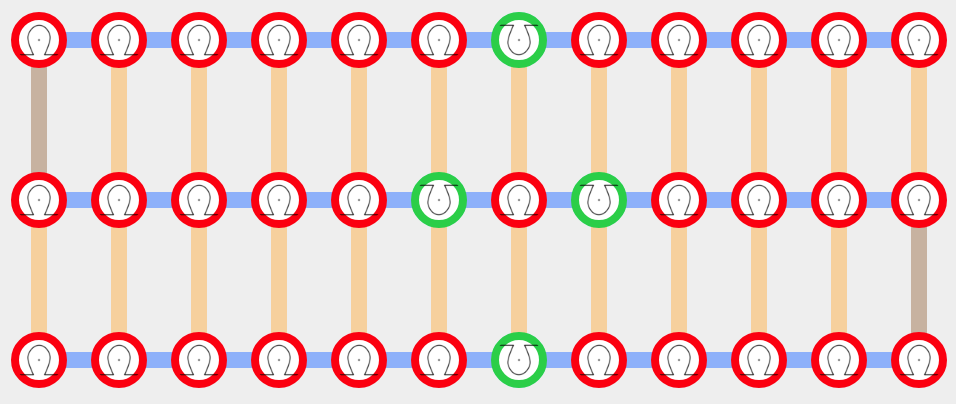}{Singular behaviour}
\subfigh{\layerHeight}{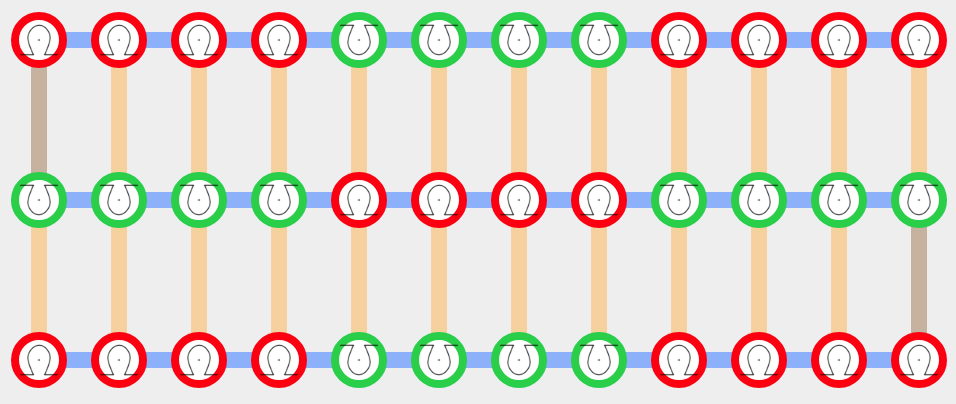}{Scalable behaviour}
\subfigh{\layerHeight}{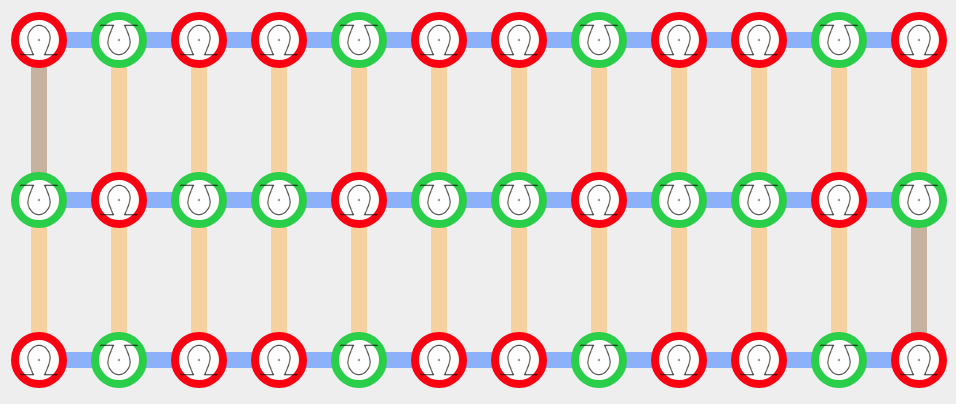}{Tileable behaviour}
\caption{
A base $3\times 3$ pattern and the different resampling behaviours of each of our layer types.
}
\label{fig:behaviours}
\end{figure*}
}

%
% Mannequin results
%

\newcommand{\figMannequin}[1][t]{
\begin{figure}[#1]
\centering
\includegraphics[width=\linewidth]{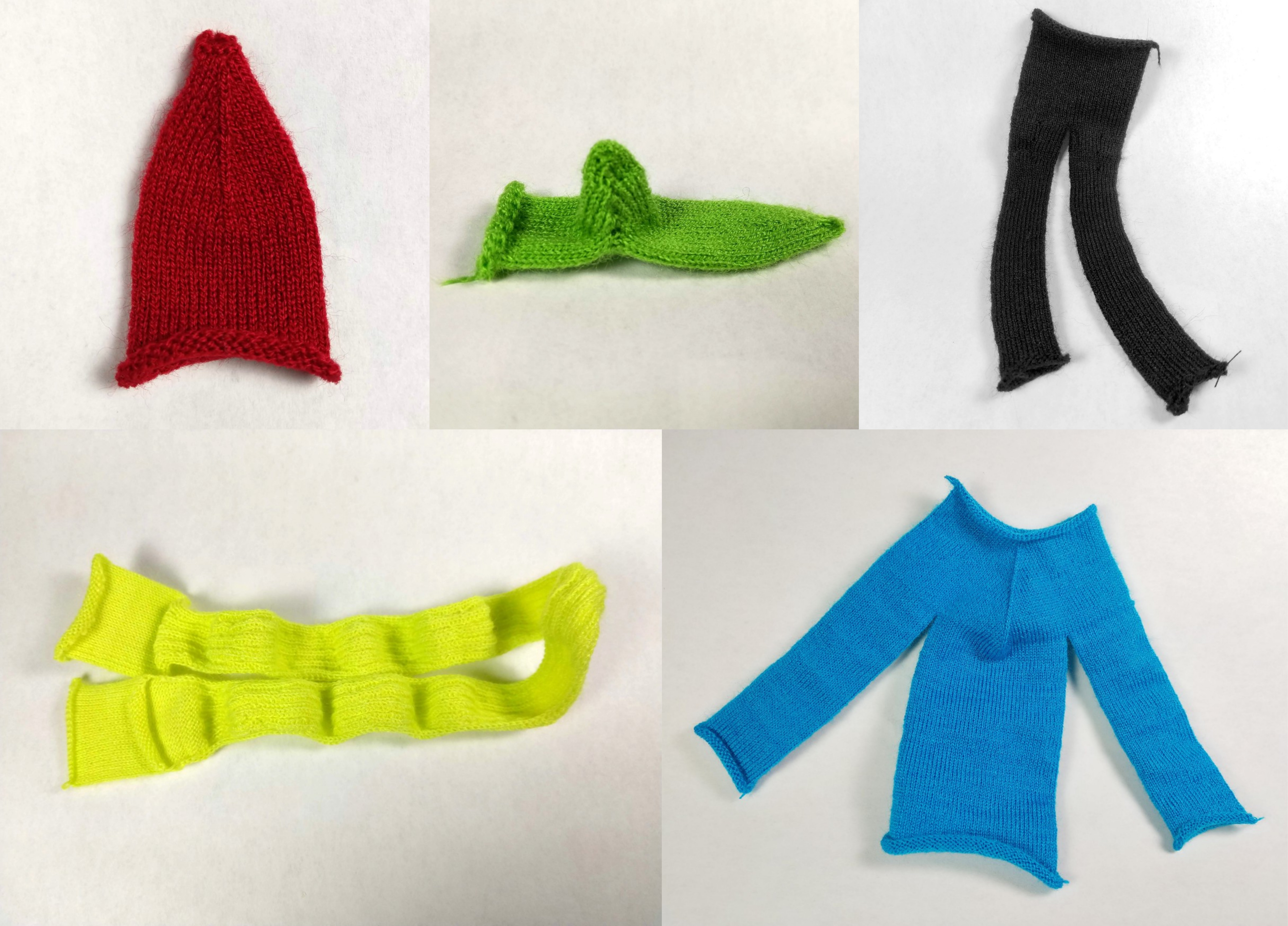}
\caption{
The individual garment pieces from Figure~\ref{fig:teaser}.
The scarf uses a single-sided part that would curl on itself by default.
Thus we used simple $2$ by $2$ ribs to keep it flat.
}
\label{fig:mannequin}
\end{figure}
}

%
% Pattern sizings
%

\newcommand{\figPatternLayers}[1][t]{
\begin{figure}[#1]
\centering
\includegraphics[width=\linewidth,trim={0 4mm 0 8mm},clip]{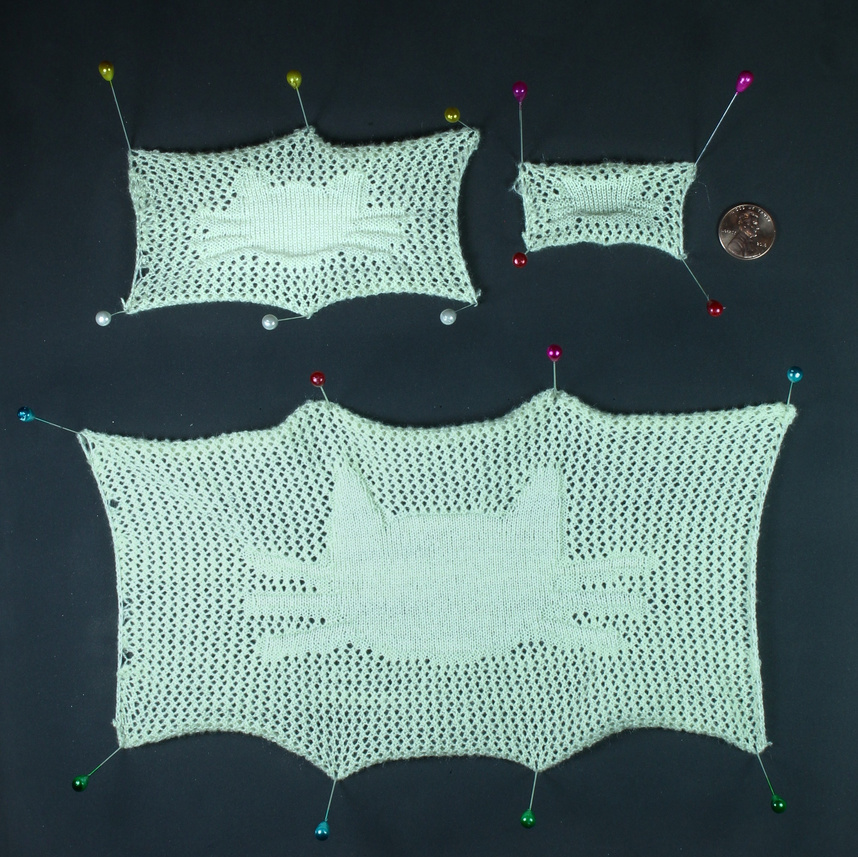}
\includegraphics[width=\linewidth]{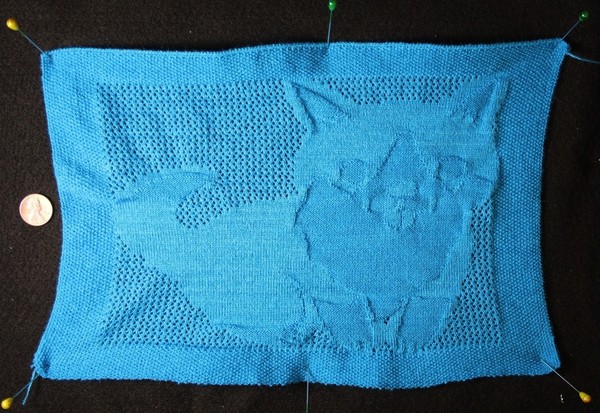}
\caption{
\textbf{Top}: impact of shape size on a two-layer pattern.
The holes are tileable moves from Figure~\ref{fig:types}.
The foreground cat is scalable and stretches with the shape.
\textbf{Bottom}: a similar four-layer pattern with an additional programmatic margin,
and two scalable foregrounds for different shades of a Corgi. 
}
\label{fig:pattern_layers}
\end{figure}
}

\newcommand{\figScarfLayer}[1][t]{
\begin{figure*}[#1]
    \centering
    \includegraphics[width=\textwidth]{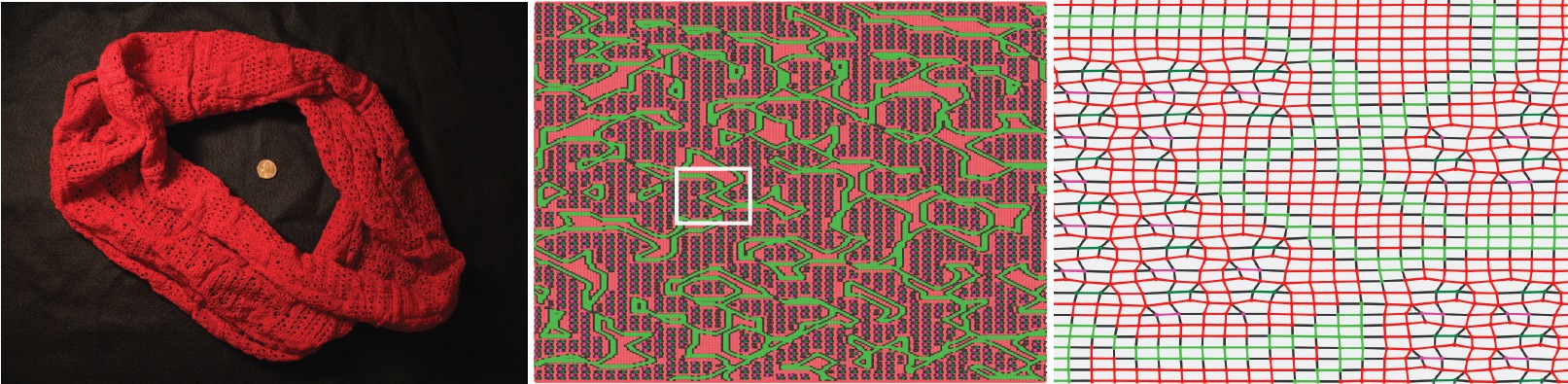}
    \caption{Visualization of the pattern of our infinity scarf (left) with our mesh visualization (center) and a close-up (right).}
    \label{fig:infinity_scarf_mesh}
\end{figure*}
}

%
% Glove patterning
%

\newcommand{\figGlove}[1][t]{
\begin{figure}[#1]
\centering
\includegraphics[width=\linewidth,trim={5mm 10mm 15mm 25mm},clip]{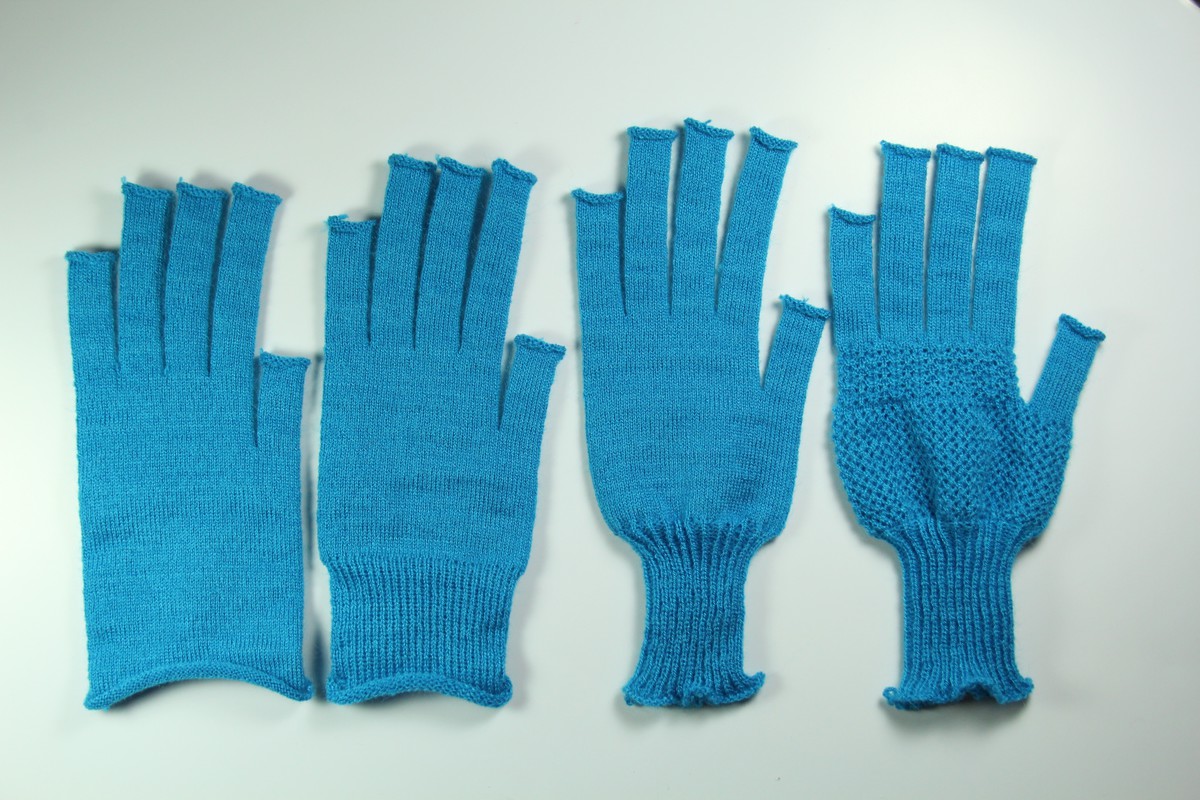}
\caption{
Patterning the glove for the teaser from left to right:
base shape,
cuff in half gauge,
half-gauge cuff with a rib pattern,
and final glove with transferred hole pattern
on main palm, as well as an additional pattern for the 4 fingers palm.
}
\label{fig:pattern_glove}
\end{figure}
}

%
% Beanie results
%

\newcommand{\figBeanies}[1][t]{
\begin{figure}[#1]
\centering
\includegraphics[width=\linewidth]{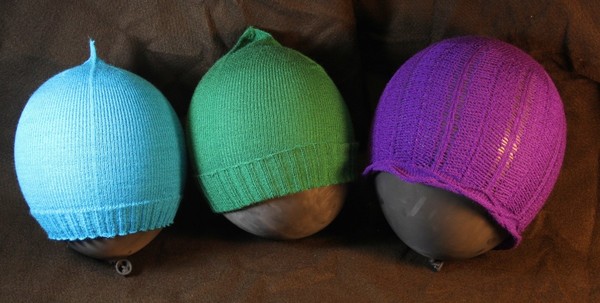}
\includegraphics[width=\linewidth]{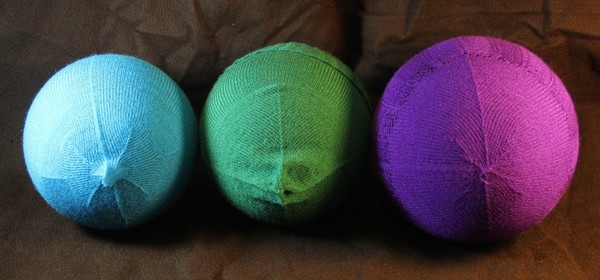}
\caption{
A reference beanie on the left and two customized beanies on its right.
The rightmost one required a few passes to adjust the lace pattern and its tension.
}
\label{fig:beanies}
\end{figure}
}

\newcommand{\figBeanieCloseup}[1][t]{
\begin{figure}[#1]
\centering
\includegraphics[width=\linewidth]{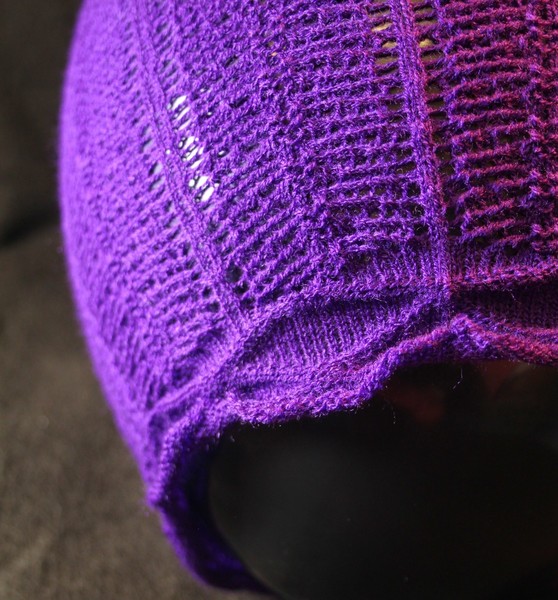}
\caption{
Beanie closeup showing the main section's lace and the curled brim with a knit/purl zigzag.
}
\label{fig:beanie_closeup}
\end{figure}
}

%
% Glove results
%

\newcommand{\figGloves}[1][t]{
\begin{figure}[#1]
\centering
\includegraphics[width=\linewidth]{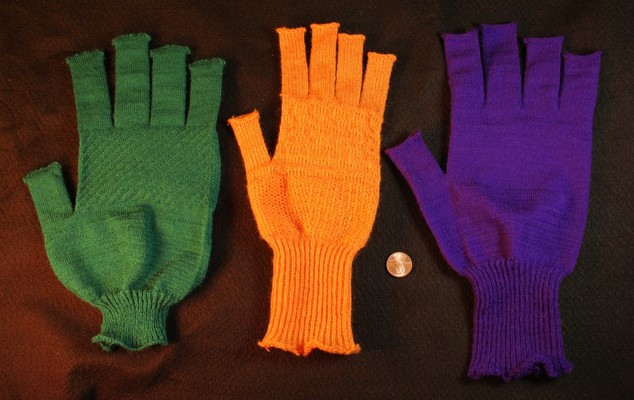}
\caption{
Three glove variants. The green one took multiple attempts because of the complicated tension requirements associated with continuous cross patterns.
}
\label{fig:user_gloves}
\end{figure}
}

\newcommand{\figBrokenGlove}[1][t]{
\begin{figure*}[#1]
\centering
\includegraphics[width=\linewidth]{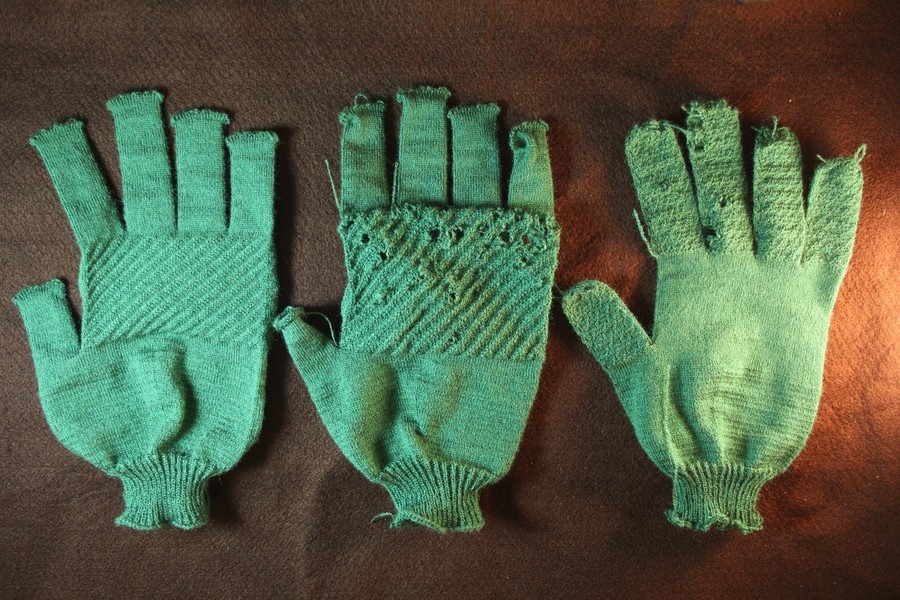}
\caption{
The evolution of the green glove from right to left.
}
\label{fig:broken_glove}
\end{figure*}
}

\figTeaser[ht!]

\begin{abstract}
 This work presents a novel interactive system for simple garment composition and surface patterning.
Our approach makes it easier for casual users to customize machine-knitted garments, while enabling more advanced users to design their own composable templates.
Our tool combines ideas from CAD software and image editing: it allows the composition of (1) parametric knitted primitives, and (2) stitch pattern layers with different resampling behaviours.
By leveraging the regularity of our primitives, our tool enables interactive customization with automated layout and real-time patterning feedback.
We show a variety of garments and patterns created with our tool, and highlight our ability to transfer shape and pattern customizations between users.
\end{abstract}

% ACM Classfication



% Author Keywords

% Print the classficiation codes

% text content
\section{Introduction}
\label{sec:introduction}

Recent advances in textile manufacturing are poised to revolutionize our everyday garments~\cite{Poupyrev16}, communication capabilities~\cite{Rein18}, and possibly even our health monitoring~\cite{Maziz17, Rein18}.
%Recent advances in textile manufacturing are poised to revolutionize our everyday garments~\cite{Poupyrev16}, communication capabilities~\cite{Rein18}, and possibly even monitor and track our health~\cite{Maziz17, Rein18}.
%Concurrently, the maker community has found various applications for textiles in additive manufacturing~\cite{Rivera17,Vogl17,Hamdan18},
The maker community has also found various applications for textiles in additive manufacturing~\cite{Rivera17,Vogl17,Hamdan18},
made ingenuous uses of programmable textiles~\cite{Albaugh19, Perner-Wilson10},
and created novel processes~\cite{Peng16}
and low-cost machines that target industrial textile manufacturing~\cite{OpenKnit, OpenKnitInstructables, Kniterate}.
This work is specifically interested in whole-garment knitting using industrial knitting machines~\cite{Shima11, Stoll11}, which provide digital control over each single loop of yarn being created, while requiring minimal user intervention during the manufacturing process.

These machines are capable of creating whole garments with full customization, but unfortunately their programming is complicated and requires skilled experts.
Low-level machine programs are typically represented as colored needle-time images (Figure \ref{fig:machinecode}) whose pixels are per-needle operations that get compiled into machine instructions.
Existing commercial softwares also provide garment templates with a set of customizable properties (width, length, etc.)
for predefined shapes (such as gloves, hats, or sweatshirts).
However, template parameters only modify the garment geometry (known as \emph{shaping});
the surface texture and appearance (known as \emph{patterning}) must be modified through separate tools, or by manipulating the local instructions generated from the template~\cite{Shima11}.

Our system unifies and simplifies these existing approaches within a single workflow, while trying to alleviate two of their original limitations:
(1) existing templates cannot be composed, which limits their potential to a fixed set of predefined shapes, and
(2) these templates lack a bidirectional decoupling between shaping and patterning, so any alteration of shape parameters requires recreating the associated patterns.

We propose an interactive system to compose parameterized knitting primitives
similar to knitting templates, while ensuring knittability and allowing continuous user customization of both garment shape and patterns.
The contributions of our work include:
\begin{itemize}
\item A web interface for creating and customizing knitting templates for everyday garments
\item A domain specific language for patterning
\item A novel layered pattern representation with different resampling behaviours for decoupling shapes and patterns
\end{itemize}

Furthermore, we release our web interface and its source code to allow the democratization of computer-aided knitting.
See the project page: \url{http://knitskel.csail.mit.edu}.
\section{Background and Related Work}
\label{sec:related}

\figStitchUnit

\figIntro

\figTNB

We begin with a brief overview of knitted fabric and the mechanisms of the knitting machine we use to create it.

The smallest unit of knitted fabric is a \emph{stitch} (as illustrated in Figure~{\ref{fig:stitch}}).
Typically, a stitch is created by pulling a yarn loop through a pre-existing stitch; this new loop is later stabilized by having another stitch pulled through it. 
The neighborhood of each stitch defines the local topology of the fabric, and must fulfill certain criteria to prevent the fabric from unravelling.
In this work, we refer to rows of stitches as \emph{courses}, and columns as \emph{wales}.
Every stitch has two course neighbors, except when the yarn starts and ends.
Regular stitches also have two wale neighbors: the preceding stitch it was pulled through, and the subsequent one that stabilizes it.
However, irregular stitches have more or fewer wale connections due to \emph{increases}, \emph{decreases}, or the yarn's start and end.

Our system is targeted at knitting garments on weft knitting machines.
In particular, we target \emph{V-bed} machines\footnote{This work uses a Shima Seiki SWG091N2 with 15-gauge needles.}, which consist of two parallel beds of stitch needles,
angled toward each other in an inverted $V$ shape.
Individual needles can hold multiple loops of yarn in their hook, as illustrated in Figure~\ref{fig:intro}.
They can be actuated as one or multiple carriers bring yarn over the beds.
The main per-needle operations include \emph{knit} which pulls a loop of yarn through the current ones,
\emph{tuck} which adds an open loop in the needle hook,
and \emph{miss} which moves the yarn without triggering the needle operation.
Furthermore, each needle on a given bed has a counterpart that is directly across the gap, on the opposing bed; these needle pairs can perform a loop \emph{transfer}, which moves a loop from one bed to the other.
Although the needle positions are fixed within each bed, the lateral offset between the two beds (known as \emph{racking}) can be manipulated. 
Thus, the counterpart of a given needle can change.
By combining racking and transfers, loops are able to move laterally along the beds.
For a detailed review of how these machines work, we refer the reader to the knitting reference by Spencer~{\cite{Spencer01}}, 
the thesis of Underwood~\cite{Underwood09},
and the algorithmic review of McCann~\etal~\cite{McCann16}.

\subsection{Knitting As A Service}

Current industrial knitting machines are expensive.
Although maker projects have attempted to create low-cost versions~\cite{OpenKnit, Kniterate}, we still would not expect users to own such machines because they require expertise to operate.
Just as manufacturing-grade 3D printing is generally not done at home but through a 3D printing service\footnote{
See \relax\url{https://www.shapeways.com/}.
}, this work follows a recent vision of \emph{knitting as a service}\footnote{
See \relax\url{https://ministryofsupply.com/}, \url{https://unless.me/}.
},
in which users send their knitting jobs to a company, which then ships back their customized knitted product.
This requires a means for the user to specify their custom-made garment, which motivates our work.

\subsection{Knitting 3D Meshes}

Several early works have tackled the problem of converting a 3D mesh to hand knitting instructions such as for plushes~\cite{Igarashi08a, Igarashi08b}.
More recently, Narayanan~\etal~\cite{Narayanan18} tackled the automatic machine knitting of 3D meshes, and posited a theoretical description of \emph{machine knittable} structures
\footnote{Note that \emph{machine knittability}, which we require, is more constraining than \emph{hand knittability}.}.
This description motivates our shape skeleton.

We avoid the external 3D modeling step, by building a CAD system for whole garment knitting in the traditional sense, where we assume modeling is done from scratch.
The first motivation is that the primitives for garment knitting are less various than the diversity of topologies that can be represented by general 3D meshes.
The second reason is that surface patterning requires specifying knitting patterns on top of the shape.
Texture mapping could allow this on meshes, but it introduces several unnecessary challenges~\cite{Yuksel19} and tightly couples the pattern with the shape.
Instead, we seek to decouple shape from pattern; thus, we use a novel layered pattern representation for general graphs.

\subsection{Shaping Primitives}
McCann~\emph{et al.}~{\cite{McCann16}} built a similar CAD-like approach, with a language and algorithm for the machine knitting of general tubular shapes.
One of their contributions was the time-needle bed visualization (see Figure~\ref{fig:tnb}), which we build upon.
They also introduced machine knitting building blocks (e.g. sheets, tubes, short rows), which are manually instantiated onto the time-needle bed.
Our system uses similar building blocks, but provides automatic bed layout.
Additionally, we tackle the problem of surface pattern customization while allowing the continuous editing of both shape and pattern.

\subsection{Modeling Garments and Patterns}

Multiple works have looked at modeling the garment making process.
Sensitive Couture~\cite{Umetani11} tackles the specification of sewing patterns for woven structures.
Other works focus on modeling the yarn directly on top of specialized structures such as \emph{stitch meshes}, with applications in realistic rendering \cite{Kaldor11,Yuksel12}.
More recently, these have been translated to \emph{hand knittable} structures~\cite{Wu18a, Wu18b}.
We build upon a similar stitch unit as theirs, but do not use 3D meshes.
Instead, we use a multi-layered 2D embedding that enforces \emph{machine knittability} and allows pattern design similarly to image editing.
This allows us to reuse the plethora of manually-codified pattern recipes~\cite{Donohue15, Butterick16}, as well as those directly inferred from images~\cite{Kaspar19}.

\subsection{Parametric Design}

Our system takes inspiration from traditional CAD softwares, together with novel systems that they have inspired. One example is Robogami~\cite{Schulz17}, which combines custom-made robot components for easy fabrication.
Similarly to the shape- and volume-authoring tools Antimony \cite{Keeter13} and Foundry~\cite{Vidimce16}, we use composable primitives, and a pattern representation based on short function expressions that can be modified by the user.

\subsection{Domain Specific Language}

Various \emph{Domain Specific Languages} have been designed to tackle dedicated problems,
such as shade trees~\cite{Cook84} for rendering,
or Lindenmayer systems~\cite{Prusinkiewicz12} for generating complex structures.
More recently, the Knitout language~{\cite{McCann16}} proposes a set of low-level instructions and a compiler dedicated to making machine knitting independent from the machine target.
Another recent set of instructions~{\cite{Kaspar19}} targets machine knitting of patterns.
We introduce a patterning DSL that employs this last instruction set within a two-stage process (select + apply) inspired by the dataset visualization language, D3.js~\cite{Bostock11}.
\section{Knitting Skeleton Overview} % Overview?
\label{sec:workflow}

In this section, we present a typical workflow session, and then elaborate on several individual components and features of our system. The following sections detail the shaping primitives and patterning DSL, before discussing our user feedback and results.
For interactive sessions, please see the project page.

\subsection{Typical Workflow}

Our user starts from a base shape primitive (flat or tubular sheet) 
and modifies its shape parameters (e.g. size, layout, seams) by interactively manipulating them on the time-needle bed (Figure~\ref{fig:tnb}, detailed below).
These interactions include dragging primitive boundaries for sizing, as well as dragging layout elements to change their location.

The user can also use a contextual menu to directly edit all exposed properties.
In the global context (no shape selected), users can create new shapes and define user parameters (such as a glove's base finger length).
While hovering over a shape primitive, users can rename it, delete it, or edit any of its properties.
Finally, by clicking on a shape boundary (called an \emph{interface}), the user can "extend" the given shape by connecting it to an available interface with valid matching parameters, or by creating a new primitive (which will automatically connect to the selected interface, and assume matching parameters).

After creating the desired shape, the user switches to the pattern editing mode, where the toolbar actions affect individual stitches.
In this mode, the user can either (1) draw the desired pattern directly onto a shape primitive, similarly to a pixel editing program, or (2) write pattern programs using our DSL in an editor.
As the user zooms in/out, we display varying levels of information, including the type of pattern operation, and the local yarn topology (course and wale connections).

Finally, the user can visualize the yarn structure with a force layout tool, save the resulting skeleton, load a new one, or inspect and export the necessary machine code.

\figSkeletonGraph

\subsection{Shape Skeleton}
The recent work of Narayanan~\etal~\cite{Narayanan18} showed that
any shape whose skeleton can be drawn on a plane without self-intersection can be machine knitted.
This motivates our underlying shape representation, which is a skeleton graph whose nodes are shape primitives, and edges are connections between node interfaces, as illustrated in Figure~\ref{fig:skeleton}.
The garment shape is defined by the node types, connections and parameters.
The final surface pattern is defined by pattern layers associated with each node, and applied on the stitches locally. Together, these produce the final knitted structure.

By construction, our shape primitives allow for a continuous yarn path within each node and across interfaces, thus ensuring knittable skeletons.
However, issues can still arise since (1) shape parameters across interfaces may be in conflict (e.g. different widths), and (2) user patterns may produce unsound structures or put excessive stress on the yarn.
We identify such problems, but we do not fix them, because there is typically no ``correct'' solution without knowing the user's intention.
Instead, we issue warnings (detailed later), and let the resolution to the user.

\subsection{Time-Needle Bed}
The main visualization onto which the shape skeleton is composed is the \emph{time-needle bed}.  
This is a common representation for machine knitting~\cite{Shima11, McCann16}, illustrated in Figure~\ref{fig:tnb}.
The actual bed layout is automatically computed as the user extends or modifies the underlying skeleton.
This representation has two advantages:
(1) it directly shows the time process followed by the knitting machine, which allows us to produce interpretable warnings if the user creates undesirable knitting structures (see Figure~\ref{fig:warnings}), and
(2) it introduces a grid regularity, which allows the user to draw complex patterns in a manner similar to layered image editing.

\subsection{Yarn Interpretation and Simulation}
Our system interprets the yarn path through time to provide warnings and errors to the user as they create their shape and combine patterns, as shown in Figure~\ref{fig:warnings}.
The main issues we catch are (1) unsafe yarn tension prone to yarn breakage, (2) too many yarn loops on a needle, risking pile-up or failed operation, and (3) reverse stitches when the opposite bed is occupied (e.g. in full-gauge tubular knitting).
We provide feedback both textually with the types of issue and potential fixes, and visually by highlighting the problematic stitches together with their conflict dependencies.

\figWarnings[t]

We also provide a force-layout graph simulation~\cite{ForceGraph} as an approximate preview of the yarn deformation after knitting, as illustrated for a glove in Figure~\ref{fig:simulation}.

\figSimulation

\subsection{Low-Level Machine Code}

Finally, we provide a view to inspect the low-level code that is generated for the current layout, as illustrated in Figure~\ref{fig:machinecode}.
This allows experienced designers and machine operators to inspect the actual program used by the machine.

\figMachineCode

\section{Shaping Primitives}
\label{sec:primitives}
Each of our three knitting primitives (\emph{Sheet}, \emph{Joint}, and \emph{Split}) play a specific role in the garment's final shape.
We detail each primitive and its properties, then provide more details on our skeleton editing paradigm. 

As skeleton nodes, all primitives have a \texttt{name} (for visualization and pattern references), a \texttt{pattern}, and a \texttt{gauge}, which we detail in the next section. 
All nodes also define a set of \emph{interfaces} which can be connected to other nodes.

\subsection{Sheet / Tube}

The \emph{Sheet} primitive is the base component for knitting any flat or tubular structure.
Its two main parameters are its \texttt{length}, defining the number of courses making up the sheet, and its \texttt{width}, defined over the length.
While the default configuration is a regular rectangle (sheet) or cylinder (tube), users can create more interesting shape profiles through stitch \emph{increases} and \emph{decreases} (illustrated in Figure~\ref{fig:stitch}). These modulate the number of stitches on consecutive courses, in order to grow and/or reduce the shape over the bed.
At a high-level, the user can modulate the width as a piecewise linear function over the normalized length interval $[0;1]$, yielding non-rectangular profiles.
If desired, the user can also customize the stitch-level shaping behavior (i.e. placement of stitch increases and decreases) using one of multiple predefined behaviours, or a user-provided function that specifies how to allocate wale connections when changing the course width.
We provide details for these functions in the supplementary material, including examples of how the shaping behaviour affects the appearance of the yarn with the location of seams.
The primitive layout can be customized by choosing a specific \texttt{alignment}, which impacts both the bed layout and the yarn stress distribution.
This primitive has two interfaces: the \emph{top} and the \emph{bottom}.

\figSheet[ht!]

\subsection{Joint}
Our \emph{Joint} primitive captures the second shaping process, called \emph{short rows}, which only knit across a subsection of the current course, while suspending the other stitches.
These partial rows induce bending in the structure, as in a sock heel. 
A \emph{Joint} represents a collection of such short rows.
The user can specify the number of short \texttt{rows}, the \texttt{width} of each, and their respective \texttt{alignment}.
Users can also specify a \texttt{layout}, which controls the normalized location of short rows along the interface boundaries, i.e. their offset for flat knitting, or the rotation for tubular knitting.
The interfaces are the same as for the \emph{Sheet} primitive.

\figJoint[ht!]

\subsection{Split / Merge}
Finally, our \emph{Split} primitive allows for more complicated structures (like gloves) that require topological branching and/or merging.
It merely consists of a \emph{base} interface and a set of \emph{branch} interfaces.
It has a branching \emph{degree} together with a branch \emph{layout}.
For automatic layouts, the user can also provide the branch \texttt{alignment}.
Furthermore, for tubular bases, the branching can be \emph{folded} (tubular branches) or not (flat branches) as illustrated in Figure~{\ref{fig:folded}}. 
Flat bases only allow flat branches.
Finally, since the interface connections are not restricted in any direction,
this primitive can be used both to subdivide an interface or to merge multiple interfaces.

\figSplit[ht!]

\figFolded[t]

\subsection{Editing Primitive Parameters}
Our system allows multiple interaction strategies.
One can work exclusively with the abstract skeleton graph, and edit parameters using a tabular inspection panel.
Alternatively, one can drag the mouse to interactively extend the shapes on the bed layout.
In this approach, more complicated parameters can be specified using the contextual menu that allows the same fine-grained control as the tabular parameter panel.

When specifying parameters through the input panel or the contextual menu, the user can enter either direct numbers, or \texttt{\#expressions} that introduce global \emph{design parameters}.
These are global variables that can be reused across different inputs and node parameters, providing a means to expose important design parameters (such as a glove's width or length scale).
For example, the user could specify the length of a glove finger via \texttt{(\#Len~+~\#LenDelta)} which introduces two parameters, \texttt{\#Len} and \texttt{\#LenDelta}.
Each of these could be independently applied for the other finger specifications,
and any changes to the variable value would be reflected globally.
These expressions can also refer to node properties with \texttt{@prop}.
For example, the width of a sheet could be made equal to its length using the expression \texttt{@length}.

\section{Patterning}
\label{sec:patterns}

\figTypesShort

\tableQuery

Given a shape skeleton, our system assembles stitches for each of its nodes, and then merges stitch information at the interfaces, producing a stitch graph whose nodes are individual stitch units (see Figure~{\ref{fig:stitch}}).
%The initial stitches have their connections defined by the shape primitives (\emph{course} and \emph{wale} connections).\liane{confusing sentence}
Initially, the stitch connections (course and wale) are defined by the shape primitives.
Each stitch also includes a pattern \emph{operation} that describes how to modify the stitch with respect to its surrounding neighborhood.
These operations allow the user to design special surface textures and appearance on top of the shape.

\subsection{Pattern Operations}

For the pattern operations of each stitch, we use the instruction set of Kaspar~\emph{et al.}~{\cite{Kaspar19}}.
The main difference is that we apply those instructions not on a one-sided regular grid, but on the stitch graph, which is then projected back to the time-needle bed to generate the final low-level machine code.
Figure~{\ref{fig:types}} illustrates the types of pattern operations we support.

Importantly, our pattern operations do not create or delete stitch units.
Instead, they modify how individual units are interpreted, 
either by providing a different operation to the target needle (\emph{purl} or \emph{tuck} instead of the default \emph{knit}),
or by altering the wale connections (\emph{miss}, \emph{move} and \emph{cross}).
In the case of \emph{miss}, any previous wale connections of the missed stitch are transferred to the subsequent one.
\emph{Move} and \emph{cross} operations change the subsequent wale connection to a neighboring one along the next course.

If a neighboring wale target does not exist (e.g. at the border of a flat sheet), then the operation is not applied.
Note that courses of tubular sheets are treated as cyclic, so a neighboring wale connection always exists (possibly on the other bed).
The system ignores \emph{move} and \emph{cross} operations on irregular stitches (increase/decrease) to ensure structural soundness.

\subsection{Patterning DSL}

To apply pattern operations on the stitch graph, we designed a \emph{domain specific language} in which the user first specifies a subset of stitches of interest -- the \textbf{query} -- onto which they can \textbf{apply} a given patterning operation.

Our types of \emph{queries} are listed in Table~\ref{table:query}, and a subset is illustrated in Figure~\ref{fig:query}.
The main query is \texttt{filter}, on which all other queries are based (with some specialized implementations to improve speed).

\figQuery

\figBehaviours

\subsection{Drawing Layers}

All our patterns are synthesized using our DSL.
We split the pattern specification into a sequence of layers:
(1) an initial global layer spanning all stitches,
(2) varying sequences of per-node layers modifying stitches of specific nodes, and
(3) a final global layer.
By default, all base layers are empty and we assume the base pattern is a standard knit stitch.

Users can write their own program, 
use pre-existing ones, or interactively draw node pattern layers in a manner similar to image editing softwares that allow pixel-level modifications.
We preview the impact of the patterns on the wales, as illustrated in the right of Figures~\ref{fig:tnb} and \ref{fig:warnings}, where \textit{move} operations displace the upper wale targets.

Our pattern layers can be exported, imported and resampled for different shapes and sizes.
The resampling behaviour can be specified by using different types of layers.
We provide three pattern drawing types -- \textbf{singular}, \textbf{scalable} and \textbf{tileable} --
whose behaviours are illustrated in Figure~\ref{fig:behaviours}.

\subsubsection{Singular Layers}
This is our default drawing mode, which does not resample the initial pattern, but simply modifies its location to account for the change in size (e.g., by centering the original pattern).

\subsubsection{Scalable Layers}
These layers resample their pattern by nearest neighbor resizing.
In this mode, we do not allow \emph{cross} operations, which are coupled in paired groups and typically applied with a limited local range constraint to prevent yarn breakage.

\subsubsection{Tileable Layers}
These layers resample their pattern by applying a modulo operation so as to create a tiling of the original pattern.

\subsubsection{From Drawings to Programs}
Drawings are stored as semi-regular grids of operations, which can be empty (for no operation, the default).
To apply the drawing, we transform it into a basic program that makes use of the drawing data together with a resampling function depending on the type of layer.
\emph{Singular} layers relocate the drawing information, whereas \emph{scalable} and \emph{tileable} layers use the \texttt{stretch} and \texttt{tile} functions, respectively.

\subsection{Half-Gauge Knitting}

For each primitive, the user can choose a desired \texttt{gauge} (either \emph{full} or \emph{half}).
This property is important because
pattern operations that modify the side of the stitch operation (regular vs reverse, or ``purl'') can only occur on a given stitch if the needle directly across from it (on the opposite bed) is empty.
In the case of tubular fabric, the opposite bed holds the other side, which can lead to conflicts.
In such case, \emph{half-gauge} knitting is a typical solution, which consists in knitting on every other needle, 
with both sides offset such that any needle's counterpart on the opposite bed is empty, as shown in Figure~{\ref{fig:gauge}}.
Note that the need for half-gauge depends on both the shape \emph{and the pattern}.
Even if it is unnecessary, it may still be desirable because it creates looser fabric. 
Thus, we do not automatically choose which gauge to use, but let the decision to the user.
For full-gauge primitives, we detect conflicting patterns and show a warning to suggest switching gauge.

\figGauge[t]

\section{Results}
\label{sec:results}

Our first results explore the range of garments that are machine-knittable using our shaping primitives.
We then consider the use of our different pattern layers, and their behaviors as node parameters change.

\subsection{Span of Shaping Primitives}

Our first collection of knitted results is illustrated on the 12 inch mannequin shown in Figure~\ref{fig:teaser}, together with adult-size versions of a patterned infinity scarf and a ribbed sock.
We show the individual pieces in Figure~\ref{fig:mannequin}, each knitted with a different yarn color.
This includes:
\begin{itemize}
    \item a \emph{hat} using one cylindrical sheet with a narrow closed top and a wide open bottom;
    \item a \emph{scarf} with pocket ends, using one main flat sheet and ends that are split-converted into cylinders (one end is open to let the other end through, and the other is closed as a pocket);
    \item a \emph{yoked shirt} using one open tube split into three tubes (two for the sleeves, one for the main body);
    \item \emph{sweatpants} as a waist tube with a split branching into two tubular structures;
    \item two \emph{socks} using a joint for the heel and two tubes, one of which narrows down to the toes where it closes.
\end{itemize}

\renewcommand\textfraction{.1}

\figMannequin

\figScarfLayer

\subsection{Pattern Layers in Action}

The top part of Figure~\ref{fig:pattern_layers} illustrates how shape modifications alter two overlapping patterning layers.
The background consists of a tileable layer repeating a sequence of left and right \emph{moves} to create lacy holes,
whereas the foreground is a scalable image mask applying normal knit stitches.
As the sheet size changes, the layers behave differently: the background keeps tiling the same small move sequence, whereas the cat foreground expands with the size.
The bottom part illustrates an extension that includes a pure program layer for the margin, and two complementary scalable foregrounds layers with different stitch operations (knit and purls).

\figPatternLayers[t!]

In Figure~\ref{fig:infinity_scarf_mesh}, we visualize the mesh of a patterned infinity scarf, which uses a layer decomposition similar to the tiled lace.
However, the tiled lace is applied within a program mask that makes use of \emph{2D simplex noise} to create a random area selection.
Its boundaries are mapped to regular stitches within 2 stitches, and to purls from 3 to 4 stitches away.

Then in Figure~\ref{fig:pattern_glove} we transfer the global hole pattern to two nodes of a glove skeleton with open fingers.
This figure also illustrates the potential impact on shaping that some patterns have.
Here, the cuff of the glove has a constant width that matches the palm node.
The ribbed cuff shrinks considerably even though it is knitted over the same bed width (in half gauge).
This is the same glove as in Figure~\ref{fig:teaser}.

\figGlove[t!]

\section{User Experience}

To verify that our interface could be used by non-expert users and receive important feedback, we asked two potential users without prior knitting experience to use our system.

\subsection{Procedure}

The users were provided a $30$ minute introduction to the basic operations involved in machine knitting, including the notions of shaping and the types of stitches.
We also supplied a few sample videos of expert user sessions, and an introductory document for our user interface.
We asked them to complete a few tasks: the first ones about patterning, and the later one for shape and pattern customization.

\subsubsection{Patterning Task}

For the first subtask, the users were given a base skeleton with a single flat sheet, featuring an initial program pattern that flattens its margins (to prevent the single-sided fabric from curling up).
Their task was to draw some additional patterns on the sheet.

In the second subtask, we provided examples of lace patterns, and asked the users to create their own lace involving at least a few move operations.

For the last subtask,  we provided a more complicated template of a wristband, which is similar to the pocketed scarf in \ref{fig:mannequin}. The users had to import a pattern from their previous subtasks and apply it on the main part of the wristband. Please see the supplementary material for these results. 

\subsubsection{Shaping / Patterning Customization Task}

The second task was to customize an adult-size garment of their choice,
given an initial skeleton.

Our users chose to customize a beanie and a glove.
For the beanie, users had to change the shape profile of the top section, and optionally modify the pattern of the brim or core sections.
For the glove, the goal was to change global, shared design parameters (finger length and width, and/or cuff length) instead of directly changing the individual node widths.
The corresponding results are shown in Figures~\ref{fig:beanies} and \ref{fig:user_gloves}.
Figure~\ref{fig:beanie_closeup} shows a closeup on the laced beanie and its brim.

\figBeanies

\figBeanieCloseup

\figGloves

\subsection{Feedback and Results}

Although none of the users had prior knitting skills, each one successfully designed sophisticated, machine-knittable structures that we were able to fabricate.
Users were surprised by their new ability to customize garments, especially larger ones (e.g., the beanie).

During the design process, users cited a mismatch between their perception of the garment size (based on our bed visualization), and size of the actual knitted result.
This suggests that, beyond local editing, it may be worth tuning the relative size of wales and courses to perceptually match that of the real yarn.
However, realistic sizing is an open problem, as current tension parameters are hand-tuned and must be adapted for complicated patterns (e.g., changing the tension impacts the garment size substantially).
It will require a better simulation that takes the yarn tension into account.

Our users generally found the patterning interface intuitive, as the image editing analogy was sufficiently familiar. Still, it was difficult for them to reason about the ultimate effect of complex lace patterns.
Eventually, they discovered that the mesh simulation was more helpful to preview the pattern impact, and made extensive use of it. They alternated between the two views (layout and mesh).
Furthermore, one user found that complex skeleton constraints could hinder pattern experimentation.
Instead, they preferred to design their patterns on separate, flat sheets, then import the design onto the final structure. These behaviours emerged organically.

Although our users were allowed to create new shapes, they did not actively try to do so in our task setup.
This suggests that non-experts would likely prefer to start from templates.
Our tool might also be valuable for professionals, designers and other expert users, but we have not validated such cases.

\subsubsection{Knittability Constraints}
Flat patterns were always knitted successfully on the first try. 
However, this was not the case for complex patterns on tubular structures, such as for the adult-size garments.

We show three result beanies, the right one having required multiple iterations to work properly.
Our machine translation had very few issues, but some patterns triggered complications during the knitting process, mainly because of fabric pile-up, which arises from non-optimal yarn tension.

In the case of the gloves, all fully knitted from start without pile-up, but we discovered that the sequence of knitting had some unexpected impact on the yarn.
Fingers are typically suspended on the bed before being knitted over to create the palm and one user decided to use complex patterns on the fingers themselves.
This led to previous fingers being excessively stretched while suspended, and the yarn locally broke.
Finding the appropriate tension was the main complication for both the patterned beanie and glove, which required a few trial-and-error attempts.
We show the evolution of the green glove of Figure~\ref{fig:user_gloves} in the supplementary material.

\section{Discussion}
\label{sec:discussion}

We first discuss some of our design decisions,
and then go over our system limitations, highlighting potential improvements.

\subsection{Scope of Our Primitives}

By composing three types of primitives (\emph{Sheet}, \emph{Joint} and \emph{Split}), our design space already spans many common garments including varieties of gloves, socks, scarves, headbands, hats, pants and sweatshirts.
The garments which we do not handle nicely are traditionally sewed garments such as pullovers with \emph{Raglan sleeves} or \emph{drop shoulders}.
The general challenge is specifying continuous interfaces that glue primitives over multiple courses.
We provide further details and an investigation of potential solutions in the supplementary material.

\subsection{Smarter Pattern Layers}

Our pattern layers are not guaranteed to behave nicely when resampled or applied to a shaped primitive.
A ``correct'' behaviour is often ill-defined: % when our automated attempt fails.
for example, at the top of a vertically striped hat, should the stripes become thinner or merge together?
Thus, it is reasonable to defer to the user in such scenarios, as we do.
A different, recently proposed strategy~\cite{Hofmann19} would be to adapt \emph{calibrated} patterns to the local context.

\subsection{Handling Multiple Yarn Carriers}

We do not yet support multiple yarns on the bed at once (e.g., for intarsia).
However we envision that such specification of the yarn can be done similarly to our patterning by additionally introducing a stack of yarn at each stitch.
The main modification would be that yarn tracing would now also involve the specification and optimization of the different yarn interactions.
This would allow not only intarsia but also functional fiber routing, yarn inlays, spacer fabric, and pockets.

\subsection{Large-Scale Interactivity}

Our system runs in a browser and can currently remain interactive with human-sized gloves, socks and beanies. Patterning or shaping full-sized sweaters or sweatpants is challenging because of their scale.
Computationally, the garments require processing a very large amount of stitches. 
Their size also presents challenges for user pattern specification, as simple pixel-based operations are insufficient.

We expect to solve the computational challenge by using hierarchical data structures that do not instantiate all stitches but only the required information (e.g., at the boundaries or where the size changes).
As for the design issue, we assume a similar hierarchical process would help.
We envision using meta-patterns on higher-level stitch representations to be instantiated for the machine.
Finally, recent patch-level pattern simulation~{\cite{Leaf18}} has shown promising interactive results.

\subsection{Machine Independence}

Currently, our system exports machine code for a set of specific machine targets (Wholegarment\textregistered~knitting machines~\cite{Shima11}).
However, our design assumptions are that of general V-bed machines.
Since our system uses very regular course structures, it should be easy to support exporting \emph{Knitout} code~\cite{McCann16}, allowing us to potentially target other V-bed machines as well.
\section{Conclusion}

We presented a novel interactive system for composing knitted primitives
that encompass a large variety of common garments.
This work also introduced a domain specific language for patterning, along with novel pattern layers that enable pattern specification akin to image editing, complete with different resampling behaviours.
Many works follow, including the support for multiple yarn carriers, which will allow more complex appearance and functional customization.
We are also interested in optimizing the interactivity using hierarchical stitch representations, and providing more realistic simulation.

\subsection*{Acknowledgments}
We would like to thank James Minor for the help with the paper and voice-over,
Tom Buehler for the video composition,
our external users Tim Erps and Mike Foshey for their time and precious feedback,
as well as Jim McCann and the members of the Textiles Lab at Carnegie Mellon University for providing us with the necessary code to programmatically work with our industrial knitting machine.
This work is supported by the National Science Foundation, under grant CMMI-1644558,  grant IIS-1409310 and Graduate Research Fellowship 1122374.

\bibliographystyle{SIGCHI-Reference-Format}
\bibliography{paper}

\clearpage

\begin{center}
\Large \bf -- Supplementary for --\\
Knitting Skeletons: A Computer-Aided Design Tool for Shaping and Patterning of Knitted Garments
\end{center}

\section{Content}

The supplementary material contains:
\begin{itemize}
    % \item A video demonstrating our interactive system
    \item Implementation details regarding the process from our skeleton to machine layout and low-level program
    \item Implementation details regarding our shaping functions for specifying the local behaviour of increases and decreases in \emph{Sheet} primitives
    \item Details about the limitations of our current primitives
    \item Details about our system's performance
    \item Additional results from the non-expert user sessions
    \item Details about the making of the green glove
\end{itemize}

\figShapers

\section{Implementation Overview}
\label{sec:implementation}

We provide a brief overview of our pipeline implementation.
Code will be made available publicly.
It borrows ideas from both Stitch Meshes~\cite{Yuksel12,Wu18a} and Automatic 3D Machine Knitting~\cite{McCann16,Narayanan18}.
The initial input is the user-generated skeleton as well as a starting interface to knit from.
The stages are, in order:
\begin{enumerate}
    \item \textbf{Shape generation}: each node is translated into its individual set of stitch courses, initial wale connections are created (but not course connections across courses), and interface courses are merged across shapes;
    \item \textbf{Course scheduling}: connected components are grouped together, topologically sorted and then a course-by-course schedule is generated;
    \item \textbf{Yarn tracing}: continuity courses are generated, course connections are generated between distinct courses, creating the actual yarn path;
    \item \textbf{Pattern application}: pattern layers are translated into their corresponding programs and executed, effectively assigning a pattern instruction to each stitch;
    \item \textbf{Layout optimization}: the node courses are organized into layout groups, whose offsets and bed sides are optimized iteratively to reduce the stress between stitches;
    \item \textbf{Knitting interpretation}: the whole time-needle bed is generated, and the yarn path interpreted using both associated pattern instructions and wale connections to generate sequential per-bed passes (namely cast-on, actions, transfers, cast-off);
    \item \textbf{Knitting simulation}: the bed passes are processed to evaluate potential dangerous yarn situation (large moves, yarn piling up) and generate appropriate warning or error messages;
    \item \textbf{Bed Compaction}: for visualization purpose, we use an additional pass in which we compact the knitting bed by removing some of the suspended stitch beds
    \item \textbf{Code generation}: the bed passes are translated into low-level machine instructions including specialized needle cast-on and cast-off procedures to prevent yarn from unravelling.
\end{enumerate}

\subsection{Shape Generation}

Each skeleton node is individually transformed into a generic shape made of a sequence of stitch courses, some of which are annotated as interfaces (with respective names similar to that of the skeleton nodes).
Each shape's course is assembled with its neighboring courses using the node's shaper program, layout and alignment properties.

Then all node's interfaces are processed: connected ones are bound across shapes, and disconnected ones are either left as-is or closed if chosen by the user.

At this stage, note that some stitch connections have not been generated.
Notably, course connections across different courses require orientation information which depends on scheduling, happening next.

\subsection{Course Scheduling}

Given the various shapes and their courses, we first groups them into connected components, and then each group is separately scheduled, course-by-course.

The scheduling is done by topologically sorting the courses according to knitting order constraints: 
we require previous branches to be knitted before allowing merge operations in \emph{split} nodes.

\subsection{Yarn Tracing}

Given the course schedule, we can now trace the path of the yarn and generate course connections.
This also involves creating additional continuity stitches so that the yarn doesn't jump between far away stitches.

\subsection{Patterning}

At this stage, we have the final stitch graph, and we use our patterning layers, transforming them into programs that assign pattern instructions to each stitch.

\subsection{Layout Optimization}

From the course schedule, we generate individual disjoint needle bed assignments.
The offsets and relative sides between courses of a same node are fixed to optimal assignments without taking other nodes into accounts, creating groups of fixed bed layouts.

Then, the bed layout groups are optimized w.r.t. each-other's offsets and relative sides.
The main trick here is that we can easily pre-compute optimal alignments by approximating the yarn stress between two full beds by pre-computing their stitch pairs, as well as their corresponding center of mass, which should typically align for the minimal stress assignment.

\subsection{Knitting Interpretation}

The time-needle bed is generated by aggregating the layout groups, and a first pass interpret each bed time, bed after bed, generating consecutive per-bed passes over time.

This generates the following (potentially empty) passes for each time-step:
\begin{itemize}
    \item \textbf{Cast-on}: empty needles are cast yarn onto (requires specialized procedures to ensure the yarn catches, which is different from typical already-cast knitting);
    \item \textbf{Actions}: per-needle actions are computed for each stitch of the current time's bed according to their number of up-going wales, their locations, the previous stitch actions, and their corresponding pattern instruction. Potential actions include: \emph{Knit}, \emph{Purl}, \emph{Tuck}, \emph{Miss}, \emph{Knit Front/Back}, \emph{Kickback Knit}, \emph{Split Knit};
    \item \textbf{Transfers}: necessary transfers are processed per-side at once, with relative orderings specified by the instruction types (e.g. Cross instructions, Stack instructions);
    \item \textbf{Cast-off}: stitches that should be cleared off the bed are cast-off using dedicated procedures to prevent unraveling.
\end{itemize}

\subsection{Knitting Simulation}

Given the interpretation, the actual knitting process is simulated to discover potential issues due to large yarn stretching or long interactions between stitch operations (e.g. knit-over-miss for large miss sections, which ends up collapsing the yarn, or several loops merging onto one, beyond the needle knitting capabilities).
This produces potential warning and error messages that the user can use to correct their layout / parameters.

\subsection{Code Generation}

The bed pass interpretations are translated into low-level machine instructions, which can then be processed by the machine (typically re-compiled there to check for further issues and assign machine parameters before knitting).

\section{Shaping}
\label{sec:shaping}

In our system, shaper programs allow the user to specify the wale binding between successive course of \emph{Sheet} primitives.
Although the main change in shape from such primitive comes from the difference in course sizes, the local wale connections have two important impacts:
\begin{itemize}
    \item They induce visual seams around the change of wale flow
    \item They impact the stitch stability (and thus can lead to knitting failures)
\end{itemize}

A shaper program describes a function which links stitch units between two consecutive course rows of sizes $M$ and $N$.
Since the order is not important as we are creating bidirectional wale connections, we assume that $M\leq N$.
The program takes as input $M$, $N$, as well as the current index $i$ within the first course ($0 \leq i \leq M$), and potentially other context parameters.
Its output is a (possibly empty) sequence of mappings between cells of the first course and cells of the second course.
For two courses of same sizes, the simplest mapping is $i \to i$.
Two common shaping programs (\emph{uniform} and \emph{center}) are illustrated in Figure~\ref{fig:shapers}.
Given the mapping from the shaping program, our system then creates corresponding \emph{wale connections} between the associated stitch units.

In comparison, systems that build upon meshes such as AutoKnit~\cite{Narayanan18} or Stitch Meshes~\cite{Yuksel12, Wu18a, Wu18b} do not directly provide user control over increases and decreases.
Instead, these are implicitly encoded in the original mesh.
Figure~\ref{fig:shapers} shows that we can use such programs to control the location of seams, which is of interest when customizing a garment.

\section{Limitations of the Shaping Primitives}

\begin{figure}[t]
\centering
\includegraphics[width=\linewidth]{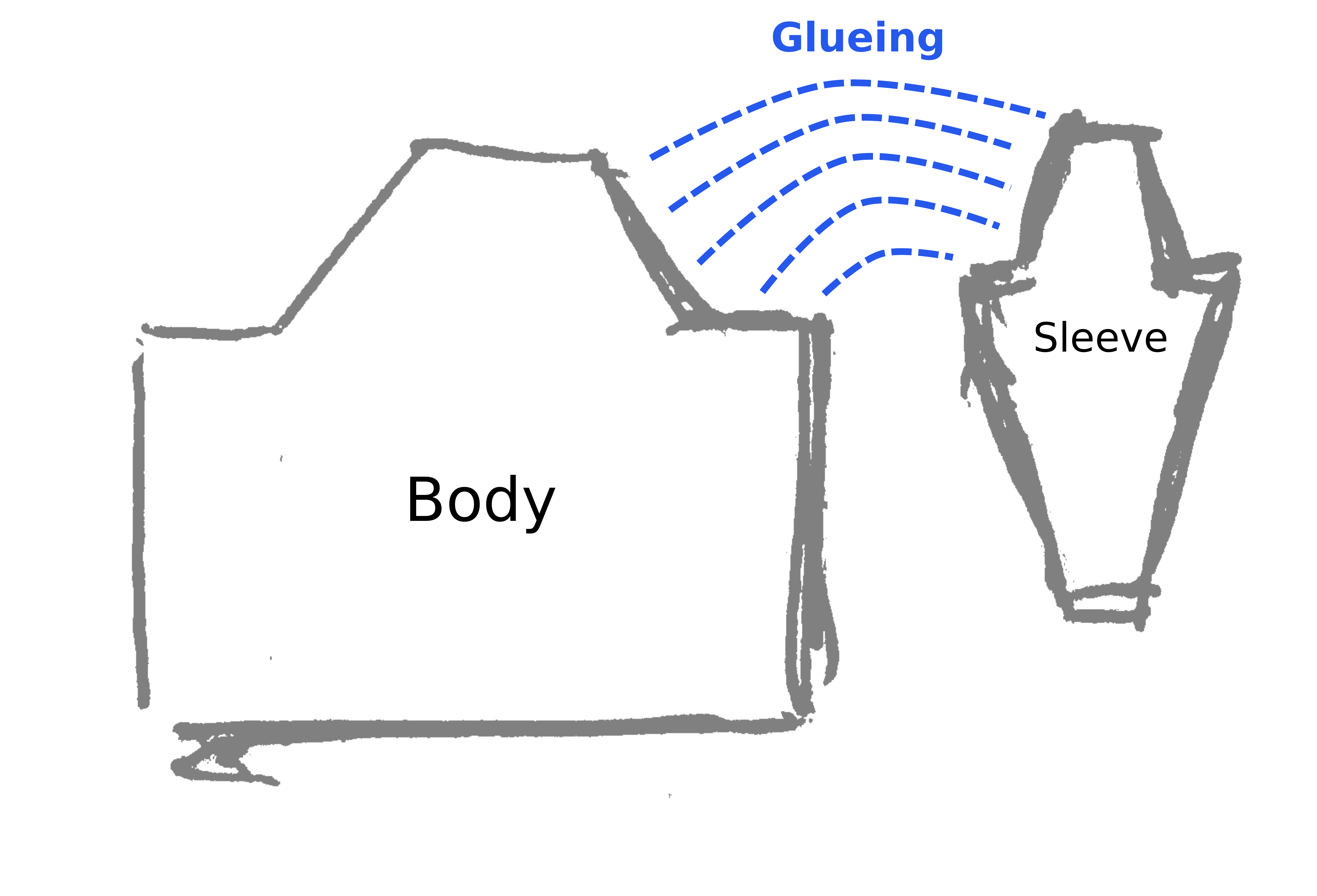}
\caption{
Illustration of one strategy to glue sleeves to the main body -- here, a Raglan sleeve.
The body and sleeves would both be knitted separately (i.e., next to each other, one at a time), and then they would be joined with a sequence of glueing operations joining both sides up to the neck section.
}
\label{fig:sleeves}
\end{figure}

One constraint of wholegarment industrial knitting machines is that they cannot instantaneously introduce a new garment section (i.e. tubular or flat sheets) with a perpendicular wale flow.
Instead, perpendicular wale flows must be joined in a continuous gluing operation that connects the two suspended sections laterally, as illustrated in Figure~\ref{fig:sleeves}.

The current primitives of this work do not support such continuous gluing operation.
This notably means that we cannot knit garments with sleeves extending perpendicularly to the base trunk.
Note that it is quite different from our \emph{Joint} primitive whose flow is continuously changed from its \texttt{bottom} to its \texttt{top}, because it does not split the flow, but only redirect it in a single direction.

Possible solutions we envision include:
\begin{itemize}
    \item Adding a specialized \textbf{T-Junction} primitive (although we are looking for a more general and simpler-to-specify primitive)
    \item Introducing \textbf{lateral interfaces} for \emph{flat} sheet primitives, which -- combined with \emph{Joint}s and \emph{Split} -- would allow the creation of \emph{T-Junction} structures (and more)
    \item Introducing an \textbf{Anchor} primitive that would allow specifying regions on top of current other primitives, to specify additional interfaces (e.g. hole sections, from which we could generate new lateral primitives).
\end{itemize}

Any of these unfortunately comes with many complication in terms of implementation and layout optimization because we cannot work with full courses alone and must allow binding courses to parts of other courses over time (i.e. for the lateral gluing operation).
There is also the problem of how the user would parameterize the gluing process as many variants exist~\cite{Budd12}.
\section{Performance}

The running time for our system is highly dependent on the client machine and web browser being used.
However, we still provide performance tables here to highlight the current processing bottlenecks:
(1) the \emph{stitch instantiation} and (2) the \emph{pattern development}.

\newcommand{\tlabel}[1]{
\multicolumn{1}{c}{\rlap{\hspace{-3mm}\rotatebox{70}{\bf #1}~}}
}
\newcommand{\thead}[2]{
%\multicolumn{#1}{c}{\vdash \hfill #2 \hfill \dashv}
\multicolumn{#1}{c}{$\vdash$ \hfill \em #2 \hfill $\dashv$}
}

\newcolumntype{b}{>{\columncolor[HTML]{F8F8F8}} c}

\begin{table*}[t!]
\centering
\resizebox{\linewidth}{!}{
  \begin{tabular}{r|bcb|cbc|b|cbc|bcb|cbc|b}
  \thead{4}{Skeleton} & \thead{3}{Shape} & \multicolumn{1}{c}{\em Pattern} & \thead{3}{Layout} & \thead{3}{Yarn} & \thead{3}{Update} & \multicolumn{1}{c}{\em Code} \\
  \multicolumn{1}{c}{\bf Name} & \tlabel{Nodes} & \tlabel{Patterns} & \tlabel{Stitches} &  \tlabel{Create} & \tlabel{Schedule} & \tlabel{Trace} & \tlabel{Develop} & \tlabel{Create} & \tlabel{Optimize} & \tlabel{Pack} & \tlabel{Interpret} & \tlabel{Simulate} & \tlabel{Compact} & \tlabel{Shape} & \tlabel{Pattern} & \tlabel{Total} & \tlabel{Generate} \\
  \midrule
cat-32x32 & 1 & 2 & 1024 & 3 & 1 & 8 & 30 & 4 & 1 & 7 & 6 & 1 & 2 & 33 & 51 & 63 & 27\\
cat-64x64 & 1 & 2 & 4096 & 8 & 1 & 15 & 21 & 3 & <1 & 18 & 14 & 1 & 5 & 65 & 62 & 86 & 40\\
glove & 10 & 3 & 5030 & 12 & 1 & 11 & 27 & 6 & 9 & 36 & 24 & 9 & 8 & 116 & 119 & 143 & 46\\
cat-128x128 & 1 & 2 & 16384 & 46 & <1 & 70 & 97 & 9 & 1 & 71 & 52 & 9 & 20 & 278 & 259 & 375 & 77\\
sock & 6 & 1 & 17740 & 46 & <1 & 50 & 89 & 18 & <1 & 70 & 46 & 8 & 34 & 272 & 265 & 361 & 101\\
corgy & 1 & 4 & 19200 & 22 & 1 & 78 & 129 & 13 & <1 & 83 & 60 & 11 & 24 & 292 & 320 & 421 & 76\\
beanie & 3 & 1 & 28774 & 141 & <1 & 79 & 124 & 29 & 1 & 107 & 67 & 7 & 51 & 482 & 386 & 606 & 96\\
noisy-scarf & 1 & 1 & 85000 & 199 & <1 & 328 & 1034 & 54 & <1 & 309 & 219 & 86 & 140 & 1335 & 1842 & 2369 & 335\\
  \bottomrule
  \end{tabular}
}
\caption{
Runtime performances of our system for the shapes within the main paper.
All times are in milliseconds.
}
\label{tab:performance}
\end{table*}

For both issues, we refer to Table~\ref{tab:performance} where we provide timings for most of the shape skeletons within the paper, excluding rendering times.
These are captured on 64-bit Ubuntu 16.04 with Intel\textregistered~Core\textsuperscript{\texttrademark} i7-3820 CPU @ 3.60GHz i7 (8 processors) and 24GB of RAM.
The test browser was Chromium 74.
To have a reasonable idea of the peak performance, we ran the profiling by loading the skeleton file, then switching to \emph{Compact} mode, and generating the output $7$ times consecutively before actually measuring the times we report here.
This warm-up leads the browser to optimize hot functions with its Just-In-Time compiler.
We then export the machine code $3$ times before measuring the code generation time.

The compaction time isn't actually needed except for shapes using branches (here mainly \texttt{glove}) but our implementation instantiates a new bed whether there is a need for it (branching) or not, so the total update times might be smaller in practice if the user does not toggle bed compaction.

\begin{figure}[t]
    \centering
    \includegraphics[width=\linewidth,trim={9mm 8mm 5mm 0},clip]{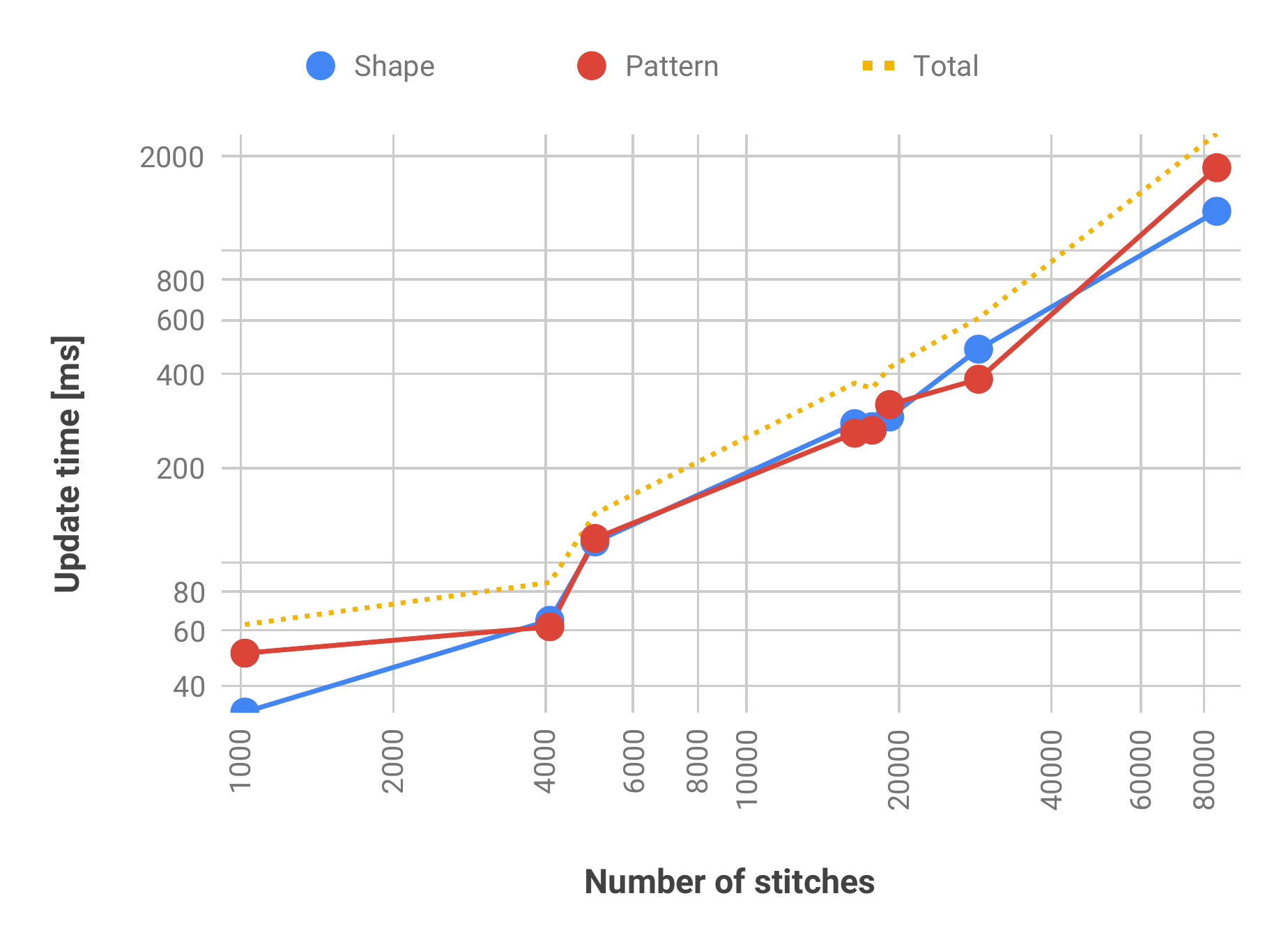}
    \caption{
    Plot of the update times for the models in the main paper.
    Shape update includes the time to rendering, minus the pattern development.
    Pattern update includes the time from pattern development to rendering.
    Both stitch and time axes have logarithmic scales.
    }
    \label{fig:performance}
\end{figure}

Generally, these timings are not meaningful as absolute numbers, but to understand the relative profile of different processes w.r.t. different shape complexities.
Both shape and pattern updates have mostly linear runtimes as shown in Figure~\ref{fig:performance} for the considered shapes.
However, the complexity of user patterns is of course going to be dependent on the user pattern.
The linear behaviour for pattern development comes from the fact that filter operations traverse all stitches, and thus most of our operations end up being linear in that number.

\subsection{Instantiating Stitches}
In the current implementation, we generate the layout from scratch every time a change happens.
This makes our implementation very simple, but prevents us from reusing most of the data that doesn't change.
Unfortunately, reusing stitch information is not easy because simple stitch modifications can have drastic impacts on the whole needle bed (e.g., adding a course at the beginning shifts all following stitches).
For the shape part, our system runs mostly linearly in the number of stitches, and thus we are limited in the size of our garment shape.

When purely editing the pattern, we avoid recreating the whole data-structure since the pattern development does not remove or add stitches.
Instead, we clear the stitch operations of all stitches, and re-apply the patterns, interpretation, simulation and compaction steps (since these are dependent on both the shape and the pattern).

\subsection{Pattern Development}
The pattern code evaluation is of course going to be longer to evaluate the more complex the pattern is.
Our DSL implementation packs all stitches in a linear array and then mainly relies on filtering the set of indexed stitches for the queries, whereas the operations are done by traversing the current selected indices and applying the operation on the corresponding stitch objects.
The implementation we provide performance for is purely CPU-based.

We expect possible improvements through a GPU implementation that would compile our DSL as shaders or other compute programs for GPGPU.
\section{Additional Non-Expert Results}

\begin{figure*}[t]
    \centering
    \includegraphics[width=0.33\textwidth]{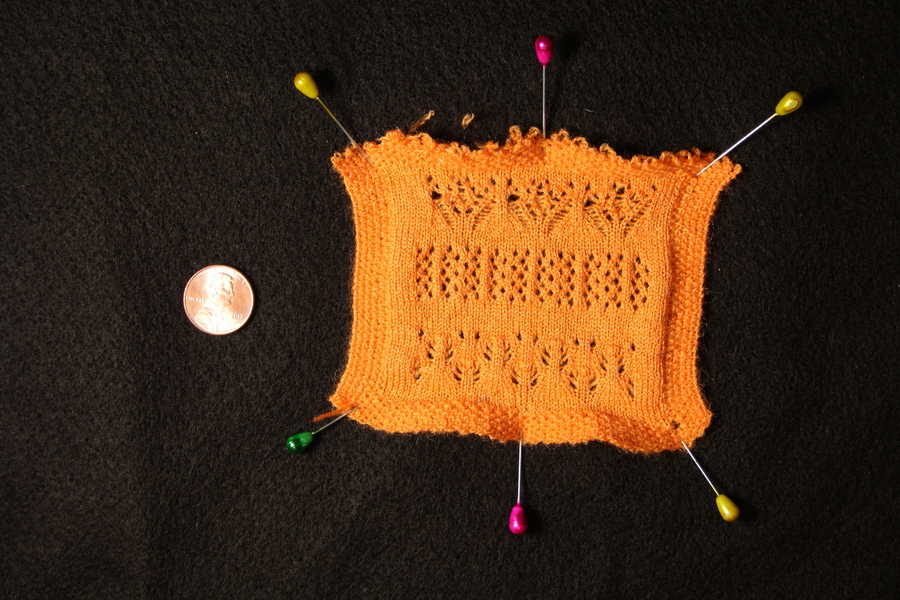}
    \includegraphics[width=0.33\textwidth]{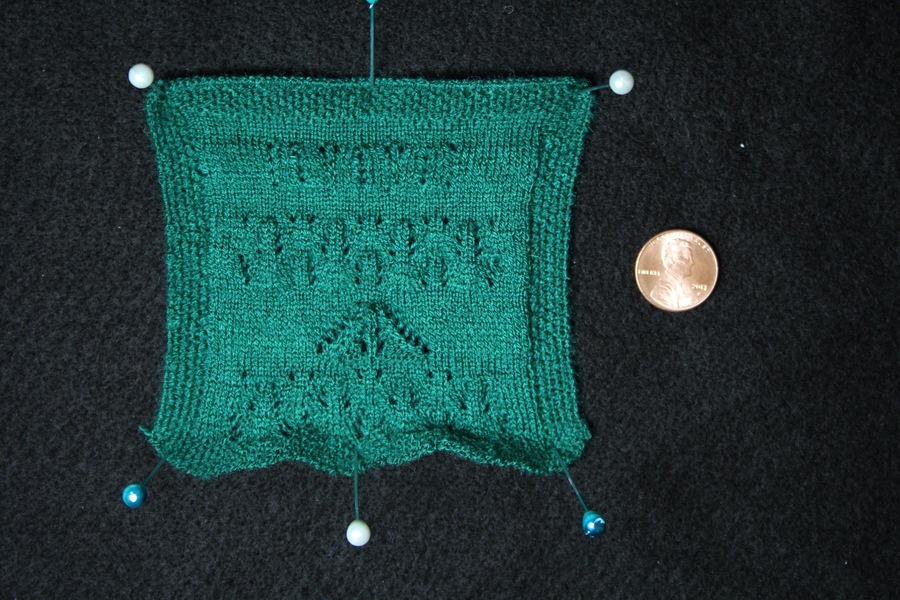}
    \includegraphics[width=0.33\textwidth]{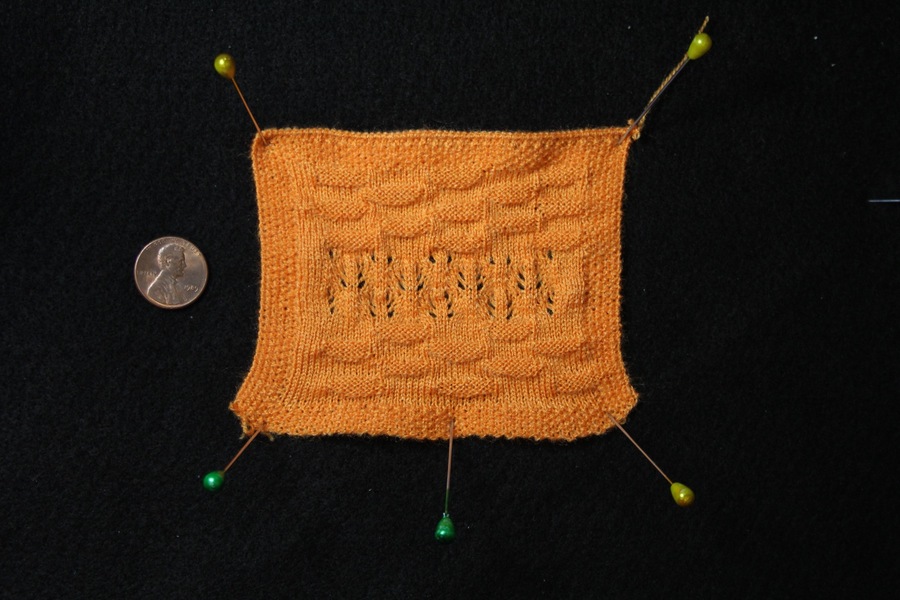}
    \caption{Lace patterns generated during our non-expert user sessions. The leftmost pattern was an expert reference that we provided for inspiration; the center and rightmost designs were novice user results.}
    \label{fig:lace}
\end{figure*}

\begin{figure*}[t]
    \centering
    \includegraphics[width=0.33\textwidth]{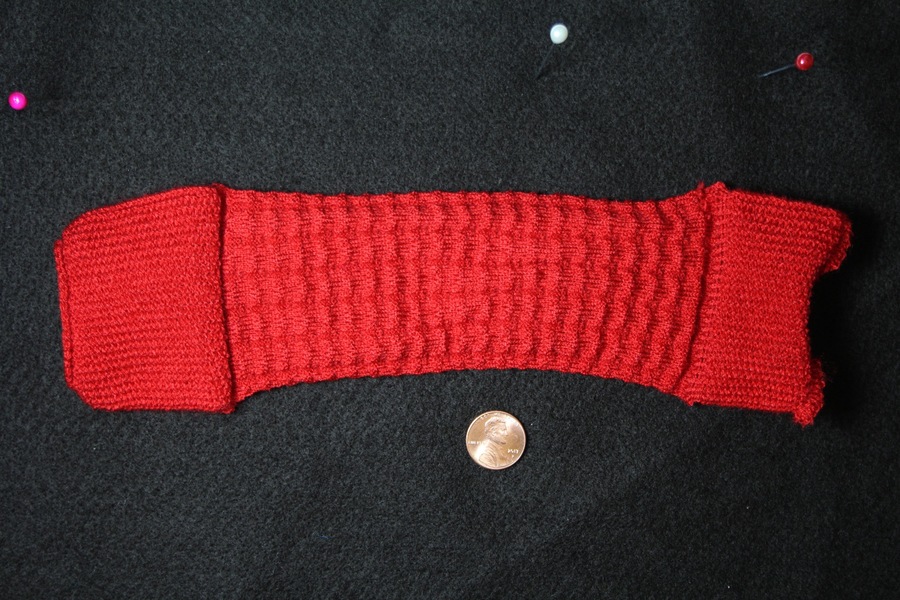}
    \includegraphics[width=0.33\textwidth]{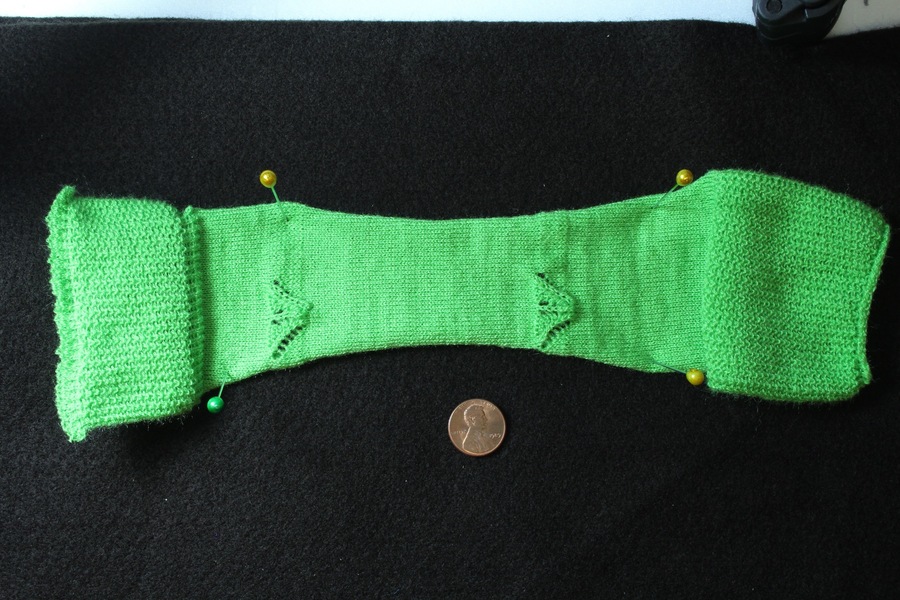}
    \includegraphics[width=0.33\textwidth]{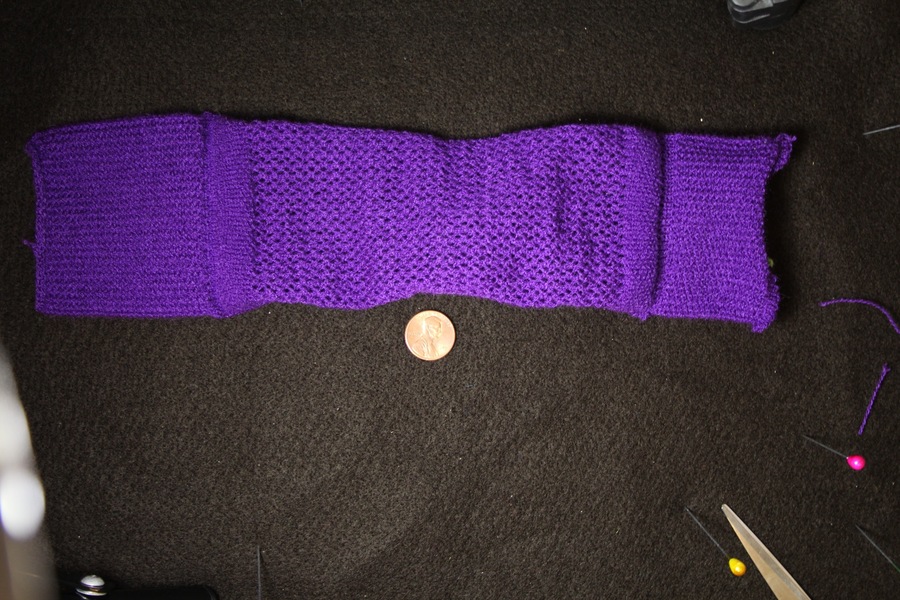}
    \includegraphics[width=0.33\textwidth]{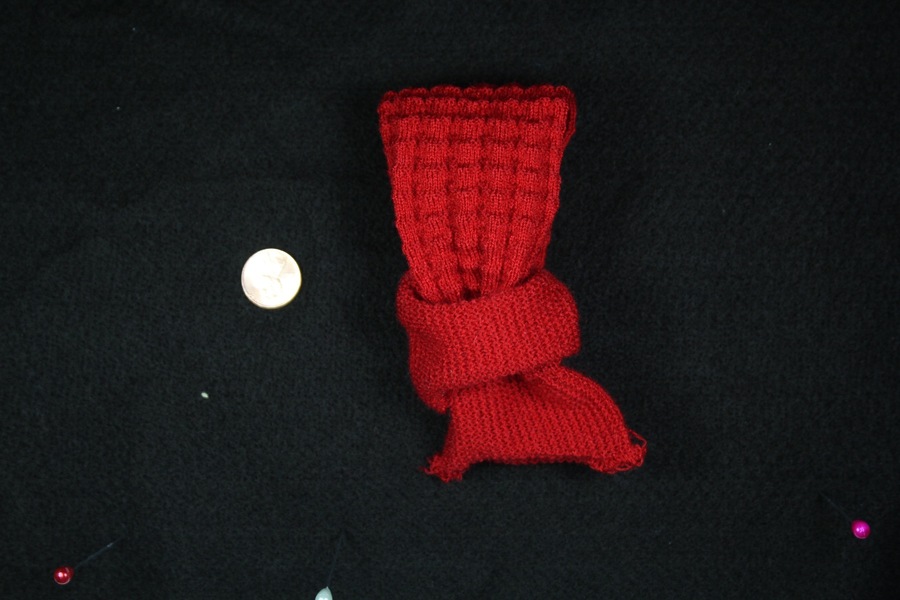}
     \includegraphics[width=0.33\textwidth]{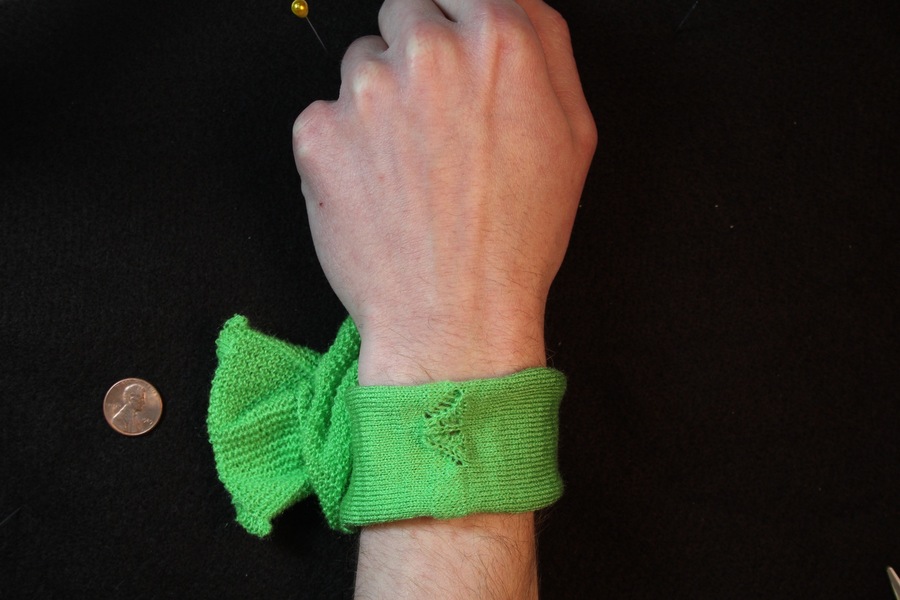}
     \includegraphics[width=0.33\textwidth]{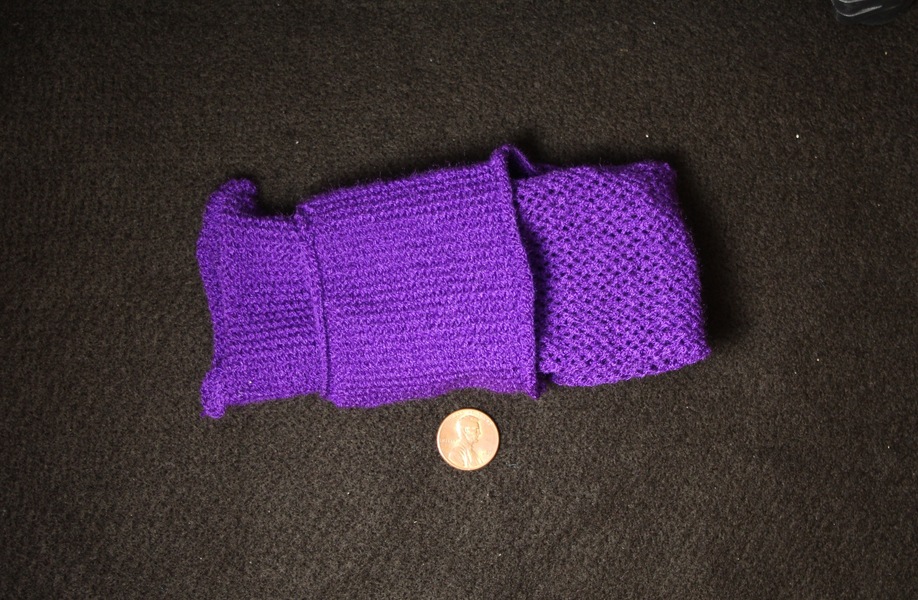}
    \caption{The third patterning task required users to transfer an existing pattern onto a provided wristband template. The patterns were either designed by users in a previous step, or selected from our repository of pre-tested designs. The leftmost column shows an expert reference; the others were designed by our users.}
    \label{fig:wristbands}
\end{figure*}

Figures \ref{fig:lace} and \ref{fig:wristbands} present additional results from the patterning task of our non-expert user sessions. Although neither of our users had previous experience with knitting, both were able to generate interesting, machine-knittable results within a few hours of their start with our system.

\figBrokenGlove

\section{Making the Green Glove}

The green glove of the last Figure from the main paper went over a few complicated iterations whose main results are shown in Figure~\ref{fig:broken_glove}.
The initial design was intending to add gripping texture under the fingers.
It was going to do so with continuously shifted cross patterns.
Furthermore, its finger ends were to be closed with some shaping at their ends.

The first issue arose from the finger ends that had used very quick increases, with unstable kickback operations next to future increases.
This led to a few holes at the finger tips.
This can be fixed by changing the shaping, but also shows that one would definitely benefit from being able to manually specify the expected shaping seams, since then one could route the yarn correctly and avoid unstable increases.
We plan to automate the detection and fixing of such problems.

The more interesting issue came from the move patterns on the fingers.
The basic glove skeleton we used was starting the knitting from the fingers to avoid having to split the fingers into multiple knitting sections.
This would have been required if starting from the cuffs, because a large knitted section is eventually suspended (the palm), from which smaller branches extends, one-by-one.
In that case, the tension at the boundary of the branches increases the farther the finger knitting is, which eventually leads to the yarn breaking.

Typically, this can easily be solved for gloves by starting the knitting from the fingers.
However, one issue is then that the fingers are kept suspended until the palm can be knitted (i.e., all fingers have been knitted).
Unfortunately, this means that patterning the individual fingers with patterns that include large movements leads to the suspended yarn being stretched continuously.
This is possible, but ended up breaking the yarn in random locations with the tension we used.

Instead, our user decided to move the pattern to the main palm, which solved the issues with the fingers (after also improving the shape of finger tips to be stable).
Unfortunately, knitting continuous cables over a large area requires a correct tension parameter.
Our initial setup was incorrectly estimating that the required yarn tension should be very loose (because of the many cables), but effectively they all combined into small $2\times 1$ shifts, that do not stretch the yarn beyond 2 needle pitches globally.
Having a wrongly loose tension typically leads to the yarn not catching correctly, and more unfortunately in it catching onto wrong needles on the sides.
This leads to unpredictable behaviours and eventually led to fabric pile up (stopping the machine and also breaking a needle during the clearing process).
Fixing the tension led to a completely fine glove.

This illustrates that although our system can allow users to create programs for shaping and patterning, there are still components missing to allow fully automated knitting as a service.
We still require some expertise in how to handle the tension and other machine parameters, which brings up many interesting avenues for research.

\end{document}